\documentclass[a4paper,12pt,letterpaper,xcolor=dvipsnames]{article}
\usepackage[margin=1in]{geometry}
\usepackage[english]{babel}
\usepackage{inputenc}
\usepackage{amsmath}
\usepackage{amssymb}
\usepackage[dvipsnames]{xcolor}

\usepackage[retainorgcmds]{IEEEtrantools}
\usepackage{graphicx}
\usepackage{tabularx}
\usepackage{makecell}   
\usepackage{multirow}
\usepackage{subfig}
\usepackage{rotating}
\usepackage{microtype} 
\usepackage{booktabs}   
\usepackage{bm}         
\usepackage{listings}   
\usepackage{verbatim}   

\usepackage[style=apa, url=true, natbib=true, backend=biber, sorting=nyt, doi=false, isbn=false, date=year, alldates=year]{biblatex}
\usepackage{url}
\DeclareFieldFormat{urldate}{}
\renewbibmacro*{url+urldate}{%
  \printfield{url}%
  \newunit\newblock
  \printfield{urldate}%
}

\usepackage{setspace}
\usepackage{hyperref}
\usepackage{acronym}
\usepackage{adjustbox}
\usepackage{titlesec}

\begin{document}

\titleformat{\section}{\normalfont\fontsize{14}{14}\bfseries}{\thesection}{1em}{} 
\titleformat{\subsection}{\normalfont\fontsize{12}{12}\bfseries}{\thesubsection}{1em}{}
\titleformat{\subsubsection}{\normalfont\fontsize{12}{12}\bfseries}{\thesubsubsection}{1em}{} 
    
\acrodef{CHE}[CHE]{Centre for Higher Education}
\acrodef{DEI}[DEI]{diversity, equity and inclusion}
\acrodef{did}[DiD]{difference-in-differences}
\acrodef{FRG}[FRG]{Federal Republic of Germany [\textit{Bundesrepublik Deutschland}]}
\acrodef{GDR}[GDR]{German Democratic Republic [\textit{Deutsche Demokratische Republik}]}
\acrodef{HHI}[HHI]{Herfindahl-Hirschman-Index}
\acrodef{OECD}[OECD]{Organisation for Economic Co-operation and Development}
\acrodef{STEM}[STEM]{Science, Technology, Engineering, and Mathematics}
\acrodef{SUTVA}[SUTVA]{stable unit treatment assumption}

\title{
    \Large\textbf{A missed opportunity? \\ Labor demand and workforce diversity
    \thanks{\footnotesize \textbf{Acknowledgments:} This research was funded by the Deutsche Forschungsgemeinschaft (DFG, German Research Foundation) under Germany's Excellence Strategy - EXC 2126/2-390838866. Boelmann gratefully acknowledges further support by the DFG through CRC TR 224 (Project A02) and by the Center for Social and Economic Behavior (C-SEB) of the University of Cologne. Bindler gratefully acknowledges further support through the DFG Research Unit ``Labour Market Transformation: Scarcity, Mismatch, and Policy'' (FOR 5675). The authors would like to thank Kathryn Herndon, Benedikt H\"{u}ls, Benedetta Muraro, Larissa Ruff, Marina Talantceva, and Joost Zickler for their excellent research assistance. We express our gratitude to Anita Krätzner-Ebert for generously sharing her expert knowledge regarding the historical context surrounding the transition of East German universities. We thank 
    discussants and participants at various seminars and conferences for their very helpful comments and discussions. All remaining errors are our own.
    \textbf{Authors:} 
    Anna Bindler - DIW Berlin, University of Potsdam and Berlin School of Economics; CEPA and RFBerlin Fellow, CEPR Affiliate. Email: abindler@diw.de.
    Barbara Boelmann - University of Cologne; RWI Essen; RFBerlin and IZA@LISER Affiliate. Email: barbara.boelmann@uni-koeln.de.
    Lena Janys - University of Konstanz; IZA@LISER Fellow. Email: lena.janys@uni-konstanz.de. 
    Luisa H. Santiago Wolf - University of Cologne; DIW Affiliate. Email: santiago-wolf@wiso.uni-koeln.de.
    }}\\
}
\vspace{-10mm}
\author{
    \begin{tabular}{ccc}
        Anna Bindler & \hspace{1cm} & Barbara Boelmann \\
        Lena Janys   & \hspace{1cm} & Luisa H. Santiago Wolf
    \end{tabular}
}
\date{{\normalsize \textbf{Last update:} \today}}
\maketitle
\vspace{-10mm}
\begin{abstract} 
\begin{footnotesize}
    {\singlespacing \noindent 
    How do labor demand shocks affect workforce diversity in the absence of targeted diversity policies? A conceptual framework illustrates the potential trade-off between the demographic and quality composition of a workforce when there is a positive labor demand shock. Exploiting the German reunification as a natural experiment, we analyze the academic labor market where nearly all social sciences professors in East Germany were replaced while STEM faculty remained largely unchanged. Using administrative data and a regional \acl{did} design, we find increased dispersion in the institutional quality of hires, indicating that the new hires came from less select departments. At the same time, female representation did not increase despite qualified women in the pipeline. Instead, East German hiring patterns converged to those in West Germany in terms of gender composition. In simulations, we investigate implied losses: Under conservative assumptions, we show that, considering the pipeline of qualified applicants, the marginal female hire's quality is approximately half a standard deviation higher than the marginal male hire's quality. 
    \\
    \textbf{Keywords:} labor demand, diversity, higher education, universities
    }
\end{footnotesize}
\end{abstract}

\onehalfspacing
\setlength{\parindent}{0pt}
\newpage


\section{Introduction}\label{sec:intro}

Workforce diversity has long been a prominent goal for governments, firms, and international organizations, motivated by both equity and efficiency considerations and reflected in various \ac{DEI} measures implemented in both the private and public sector \citep{oecd2020}.\footnote{See, for example, articles in Forbes \citep{westover2020}, the Harvard Business Review \citep{carr2019}, the Washington Post \citep{wpbrandstudio2021} or a report by McKinsey \citep{dixon-fyle2020}.} The topic gained additional importance in light of tight labor markets and the resulting discourse on diversifying the workforce to fill vacancies.\footnote{See, for example, Germany's skilled labor strategy \citep{diebundesregierung2022} or \citet{oecd2023} for an international overview of labor shortages and  migration policies.} In recent years, however, questions around diversity management have become increasingly politicized, with current developments going as far as rolling back \ac{DEI} programs entirely. For instance, the share of S\&P 500 companies with \ac{DEI} incentives for executive pay dropped from 57 (52) percent in 2023 (2024) down to 22 percent in 2025 \citep{kerber2025}. This roll-back of \ac{DEI} programs raises the question of how market forces alone will shape workforce diversification and at what cost to worker quality. Existing studies have either analyzed \ac{DEI} programs directly,\footnote{Examples include \citet{ahern2012,bertrand2019,ferrari2022,kunze2017a,maida2022,schaede2023}.} or looked at \emph{negative} labor demand shocks \citep[e.g.,][]{auer2022,berbee2025,dustmann2010}. Neither is informative about how future open positions will be filled. The key contribution of this paper is to address this gap: We go back to history to understand the extent to which \emph{positive} labor demand shocks can affect the composition of labor markets---without diversity-targeted interventions---and what that means for the (mis)allocation of talent in the labor market in particular with respect to gender.

This paper thus asks: How does the workforce composition change when labor demand expands, and what are the associated changes in workforce quality? Specifically, we study the composition---both in terms of demographics and quality---of professors in Germany when a large labor demand shock hits East German academia in the aftermath of the reunification. The setting has several features that make it especially well-suited for studying hiring under a positive labor demand shock. First, the academic labor market is characterized by a long qualification period of several years of doctoral and postdoctoral research, so that labor supply is effectively fixed in the short run and demand and supply effects can be cleanly separated. Second, the setting predates any targeted \ac{DEI} programs in German academia, allowing us to study market responses in an institutional environment free of deliberate diversity interventions. Third, administrative data allow us to observe both the pool of qualified candidates available at the time and the actual hiring outcomes, enabling us to assess whether demand shocks exhaust the pipeline efficiently.

The analysis is guided by a stylized framework in which workers can be characterized along two dimensions: a demographic characteristic and a quality signal. In this framework, hiring is subject to a trade-off between these two dimensions if the status quo workforce composition cannot be kept, e.g., in the case of a sudden increase in labor demand. We derive two scenarios for this trade-off: Prioritizing the quality dimension versus prioritizing the demographic dimension in hiring. In the first scenario, one would expect a diversification of the workforce substituting away from the predominant demographic type to preserve quality---in our case in terms of gender, increasing the share of female professors. In the second scenario, one would expect a stable demographic composition of the workforce but a reduction in average quality---in our case in terms of academic credentials, increasing the share of professors recruited from lower-ranked universities.

We investigate this trade-off empirically building on a large, unexpected, and field-specific labor demand shock in East German academia after reunification. When the East German university system, previously shaped by socialism, was reorganized after the reunification, staff deemed ideologically or productivity-wise unfit according to West-German standards were replaced. This created an acute labor demand shock in the social sciences---fields closely tied to ideological concerns, where only around 10\% of existing staff were retained---while \acf{STEM} fields were largely unaffected, with hardly any replacements taking place. We exploit this setting in a regional \acf{did} design, comparing the professors in the social sciences---the field experiencing near-total staff replacement---to those in \ac{STEM} fields---which serve as a control group---across East and West Germany. This design exploits the differential treatment across fields within East Germany, abstracting from any general East--West differences or reunification effects more broadly. We view the staff replacements as a labor demand shock in East Germany affecting primarily West German candidates, in line with our empirical evidence: Professors in the social sciences in East Germany in 1998 are younger both than their \ac{STEM} colleagues in the East and than West German professors. They mostly have a habilitation from the West, indicating that the positions opened up in the replacement process were filled by younger West Germans from the pipeline.

The estimating sample draws on administrative data covering the universe of professors in Germany in 1998, shortly after the replacement process was largely complete. These data contain information on academic field and university affiliation, year of birth, year of habilitation---the final postdoctoral qualification at the time---, gender, and, crucially, the institution at which habilitation was obtained. We use the latter as a proxy for academic quality visible to hiring committees. Departmental quality of the home institution serves as a plausible quality signal: If selection into the home institution was at least partly meritocratic, institutional reputation reflects individual quality, and a higher-reputation department may also directly enhance academic production \citep[see e.g.,][]{bethmann2023, waldinger2010}. As such, using the place of habilitation is a reasonable proxy for the quality of professors as observed by the hiring committees at the time---which is also still observable today, making it particularly valuable for our empirical analysis. 

Our analysis yields three sets of findings. First, we document that the labor demand shock translates into a decrease in average worker quality. Using the institution of habilitation to proxy for academic quality as explained above, we construct a new index of institutional dispersion among hires. This step is necessary, as a formal ranking of universities and departments is not available at the time. The intuition behind our measure is that a high concentration in habilitation institutions indicates that professors are mainly hired from select places which are perceived as high quality. Similarly, an increase in the dispersion of habilitation institutions implies that professors are also hired from lower-ranking places. We find that dispersion increased substantially for social science professors in East Germany, indicating that the demand shock was accommodated by drawing from a wider set of training institutions, rather than by recruiting more diversely along demographic lines. This increase in dispersion is disproportionately concentrated among male hires.

Second, we document that the labor demand shock did not increase diversity in the workforce, where gender is the demographic dimension of interest in our context: With the staff replacements, female representation did not increase in the social sciences in East Germany relative to the \ac{STEM} control group. This speaks against the scenario resulting form our framework which results in diversification. When compared to the differential gender gaps in the pre-reunification period, our results rather suggest that diversity \textit{decreased} in East Germany with the replacements, converging to hiring patterns prevailing in West Germany, where only 4\% of professors were female in 1988. Notably, the pipeline of qualified candidates would have supported substantially higher female shares (at least twice as large) had hiring committees drawn equally on male and female candidates. As the German reunification is a unique context, we replicate these findings in a different setting: The West German university expansion in the 1960s and 1970s increased the number of professors sevenfold from 2,745 in 1960 to 19,308 in 1988. As in the East German case, the share of hired female professors fell short of what the pipeline of qualified candidates would have permitted.

Third, we quantify the implied efficiency costs through a simulation approach. Drawing on the observed pipeline of male and female candidates, we simulate their quality distributions and identify the quality cut-off implied by the number of observed hires for each group. Under conservative assumptions of no differential selection into the pipeline by gender, we find that the marginal female hire's quality is approximately half a standard deviation higher than the marginal male hire's quality. When we instead allow for positive selection of women into the academic pipeline---a fact which is empirically well-documented \citep{iaria2024} and consistent with the many barriers women face prior to reaching professorship \citep{ceci2023}---the implied quality gap is even larger: When we assign to the best female the same quality as to the best male candidate, to the second-best female that of the second-best male candidate etc., then the worst male hire in a staff replacement position is 1.2 standard deviations lower in quality than the worst female hire. Put differently, to rationalize the observed gender-disparate hiring outcomes in a fully merit-based system, one would need to assume that female candidates are drawn from the top 15\% of the ability distribution while males are drawn from only the top 5\%.


Our paper makes three contributions. First, we contribute to the literature on labor demand shocks and the demographic composition of labor markets by looking at the so far under-researched situation of a \emph{positive} labor demand shock. Empirical evidence on the relationship of employer discrimination and labor market tightness is mixed \citep{baert2015,carlsson2018a}, while the extensive business cycle literature documents that minority employment tends to be more cyclically sensitive \citep{auer2022,berbee2025,couch2010,dustmann2010,johnston2016,hoynes2000,orrenius2010,xu2018}. The consistent finding that negative shocks hit minority workers more severely, however, does not necessarily mean that the reverse is also true, i.e., that positive shocks should increase diversity. Only few studies look at firm hiring responses to positive demand shocks: \citet{holzer2006} show that firms in the 1990s expansion placed less weight on personal characteristics in hiring, while \citet{bergman2022} show that employment-boosting expansionary monetary policy benefits under-represented demographic groups in particular in tight labor markets. \citet{illing2025} document that replacement hiring of deceased workers leads to lower wages for female replacements. Our paper offers direct evidence that a large positive demand shock need not yield more diverse hiring in the absence of deliberate policy.

Second, we contribute to the literature on the efficiency costs of workforce homogeneity and the misallocation of talent. Team performance evidence shows that homogeneous teams underperform heterogeneous ones across multiple dimensions of productivity \citep{hamilton2003,hoogendoorn2012,hoogendoorn2013} and manager diversity has been shown to increase firm performance \citep{flabbi2019}. In addition, performance costs arise when selection into positions does not reflect the underlying quality distribution.
Previous evidence on talent misallocation links under-representation to costs in economic growth \citep{hsieh2019}, firm survival \citep{weber2014}, firm profitability \citep{ashraf2024,siegel2019}, and innovation \citep{lubczyk2025,lubczyk2025a}. Our contribution here is to show that in an episode of large-scale replacement hiring---where the pool of qualified candidates is directly observable---the workforce quality losses from demographic homogeneity in hiring are substantial and quantifiable. This provides unusually direct evidence on misallocation costs, complementing studies that rely on structural identification.

Third, we contribute to a growing body of evidence on gender dynamics in high-skilled labor markets and academia specifically. Gender disparities in hiring, promotion, publication, and recognition in academia have been extensively documented \citep{ceci2014,ceci2023,iaria2024,janys2024,lariviere2013}, and our paper adds evidence on the hiring stage at the level of full professorship---a transition that has received less direct attention than earlier career stages. More broadly, our findings are likely to generalize beyond gender to other dimensions of under-representation, including socioeconomic background \citep{abramitzky2024,stansbury2023,stansbury2025} and ethnicity \citep{bayer2016,price2009}, since the mechanisms we identify---reliance on established hiring strategies and failure to exhaust diverse talent pools under demand pressure---are unlikely to be gender-specific.

\section{A Stylized Framework of Labor Demand and Worker Types}\label{sec:model}

To guide our empirical analysis in Sections~\ref{sec:workerchars} and~\ref{sec:workerdem}, we start by conceptualizing the role of labor demand shocks on workforce diversity in a stylized economic framework.
We assume that there is a pool of potential workers $P$ that consists of different worker types who differ in two dimensions. First, there are workers of characteristic $C=\{H, L\}$ which is a (noisy) quality signal (high or low quality). In the empirical analysis, we will interpret $C$ as the perceived quality of a worker. Second, workers are of demographic type $D=\left\{F, M\right\}$ which is a fixed characteristic (female and male). 

The pool of potential workers is thus $P=\{FH, MH, FL, ML\}$ where $FH$, $MH$, $FL$ and $ML$ are the potential types (females and males with high or low quality, respectively). The respective proportions of each worker type are $q_{FH}, q_{MH}, q_{FL}, q_{ML}$ such that: 
$$P = q_{FH} + q_{MH} + q_{FL} + q_{ML} \equiv 1$$ 
In the context of the academic labor market (as in our empirical analysis), $P$ can be thought of as ``the pipeline''. Depending on the shares of males and females, we label $P$ as diverse ($q_{MH}\approx q_{ML} \approx q_{FH} \approx q_{FL}$) or non-diverse (e.g., $q_{MH} \gg  q_{FH}$).

Our framework characterizes a labor market with a single skill (quality) type and no complementarities between high- and low-quality workers. Unlike models where high- and low-skill workers jointly contribute to production, here only one job type exists with workers varying in their ability to perform it well. Employers thus prefer high-quality workers.\footnote{This fits the setting of our empirical analysis: In Germany, professorial salaries are largely fixed, precluding the option to hire lower-quality professors at lower pay. This applies broadly to public-sector jobs.} Our conceptual framework matches labor markets for professors, where universities seek high-quality researchers, and extends to leadership positions more generally, such as team and firm leaders in the private sector, judges, and other public-sector managerial roles. It thus provides insight into labor demand shocks in markets for highly qualified workers.

We let the status-quo (incumbent) workforce consist of the four worker types, with respective shares defined as $\tilde{q}$:
$$ \tilde{WF} = \tilde{q}_{FH} + \tilde{q}_{MH} + \tilde{q}_{FL} + \tilde{q}_{ML} \equiv 1$$
We expect that high-quality workers are over-represented in the status quo, i.e., $\tilde{q}_{\cdot H} >  q_{\cdot H}$. This is equivalent to expecting that, overall, employers make hiring decisions in line with the organization's incentives. In addition, we assume that there is a predominant demographic type in the status quo. For simplicity, in line with gender gaps in academia, let us assume this type to be men. Thus, workers of demographic type $M$ are over-represented relative to their share in the pool of potential workers $P$: $\tilde{q}_{M\cdot} >  q_{M\cdot}$. Combining both dimensions, we thus assume that 
$(\tilde{q}_{MH} - \tilde{q}_{FH}) > (q_{MH} - q_{FH}), \hspace{4mm} \tilde{q}_{FH} > \tilde{q}_{FL} \hspace{4mm} \text{and} \hspace{4mm} \tilde{q}_{MH} > \tilde{q}_{ML}$. We assume that this status quo reflects the desired labor force composition (in terms of preferences on the demand and/or the supply side). This is plausible given initially slack labor markets where labor supply exceeds labor demand.  

Now let there be a positive labor demand shock, i.e., an exogenous increase $\Delta$ in labor demand. We denote the post-labor-demand shock composition of the resulting workforce after filling the new openings as:
$$ WF^{\Delta} = q_{MH}^{\Delta}+q_{ML}^{\Delta}+q_{FH}^{\Delta}+q_{FL}^{\Delta} \equiv 1$$

There are three scenarios to fill the new openings. 
In the first scenario, the pool of potential workers is large enough with respect to both, demographic and quality types, i.e., there is a ``filled pipeline''.\footnote{In the following, we assume that the pool of workers is always large enough overall and that positions do not remain unfilled.} In this case, the openings can be filled to keep the (desired) status quo, and thus $\tilde{q}_{\cdot\cdot} \simeq q^\Delta_{\cdot\cdot}$. In the subsequent scenarios, the pool of workers does not allow for the same labor force composition after the demand shock and the status quo cannot be kept. This results in a trade-off. Put simply, to illustrate this trade-off: If, for instance, all high-quality type men of the pool $P$ had been hired, employers could either hire low-quality type men, decreasing overall worker quality, or women of high quality, preserving (some of) the pre-shock worker quality. This idea is generalized in scenarios two and three.

In scenario two, openings are filled to keep the status-quo with respect to the demographic dimension: $q_{M\cdot}^{\Delta}=\tilde{q}_{M\cdot}$ and $q_{F\cdot}^{\Delta}=\tilde{q}_{F\cdot}$. This can happen if two conditions are met. First, if the status quo cannot be kept, the preference for keeping the demographic types stable is stronger than for keeping the quality types stable. Second, the pool of potential workers of the predominant demographic type is large enough. In this scenario, as long as demographic and quality types are not perfectly correlated, employers downgrade in worker quality (else, they would be in scenario one). Since in reality, quality is likely to be a continuous variable rather than a binary type, downgrading (with possibly more marginal adjustments) is even more likely in this scenario. Thus, we expect for workers of demographic type $d\in D$ and the predominant type $d^*$ that: 
$$ q_{d\cdot}^{\Delta} \simeq \tilde{q}_{d\cdot} \hspace{4mm} \text{and} \hspace{4mm} q_{d^*H}^{\Delta} < \tilde{q}_{d^*H}  \hspace{4mm} (\text{and thus:} \hspace{2mm} q_{\cdot H}^{\Delta} < \tilde{q}_{\cdot H})$$ 

Instead, in scenario three, openings are filled to preserve worker quality: $q_{\cdot H}^{\Delta}=\tilde{q}_{\cdot H}$ and $q_{\cdot L}^{\Delta}=\tilde{q}_{\cdot L}$. Again, this can happen if the resulting workforce composition aligns with preferences (this time for quality over demographic types) and if the pool of workers allows for it. In this third scenario, employers substitute away from the predominant demographic type $d^*$ and we expect for workers of quality type $c\in C$ that:
$$ q_{\cdot c}^{\Delta} \simeq \tilde{q}_{\cdot c} \hspace{4mm} \text{and} \hspace{4mm} q_{d^*H}^{\Delta} < \tilde{q}_{d^*H} \hspace{4mm} (\text{and thus:} \hspace{2mm} q_{d^*\cdot}^{\Delta} < \tilde{q}_{d^*\cdot})$$ 

Considering differences in the status-quo workforce $\tilde{WF}$ and the post-demand shock workforce $WF^\Delta$ allows us to study compositional changes in the workforce with respect to the scenarios defined above. More specifically, we can analyze and empirically test whether the pre-shock demographic composition remains unchanged while quality decreases (scenario two), whether the demographic composition changes while quality remains stable (scenario three), whether we observe no changes at all (scenario one) or a change in both dimensions (combination of scenario two and three).

Importantly, neither our framework nor our empirical analyses make an assumption on whether any change in the workforce composition is driven by preferences from the demand side (e.g., employers' tastes for one type) or the supply side (e.g., candidate preferences over new jobs). That means that our analysis is informative about potential changes in the workforce composition and the types of workers that fill the new demand, but cannot distinguish, for instance, between labor market discrimination (on the demand side) or mobility restrictions (on the supply side) when it comes to the mechanisms.\footnote{We note here already that (i) in terms of preferences on the demand side, hiring for the open positions was led by professors from West Germany only, and (ii) in terms of preferences on the supply side, for institutional reasons the open positions were predominantly filled with scholars from West Germany.} Our analysis is, however, useful to detect structural problems in assigning the best worker to the next job opening and to quantify their extent. This helps to understand whether (and potentially why) market forces alone may or may not be sufficient in increasing diversity among the workforce.


\section{The Reunification as a Natural Experiment}\label{sec:natexp}

We test the implications from the conceptual framework in a unique empirical setting that allows us to combine variation from a natural experiment---a large positive labor demand shock---with data on two observable worker characteristics $C$ and $D$. Specifically, we analyze the academic labor market for professors in Germany. In terms of worker characteristics, we measure (perceived) quality $C$ by academic background (``home'' university) and demographic type $D$ by gender.

\subsection{The Labor Market for Professors in Germany}\label{subsec:gerprofs}

The German tertiary education sector consists of universities, universities of applied sciences (``Fachhochschulen''), colleges of Arts and Music, theological colleges, and pedagogical colleges (``P{\"a}dagogische Hochschule'', mainly historically). In this paper, we focus on universities.\footnote{Hiring processes and labor markets for other types of tertiary education institutions are quite different.} Under the umbrella of a unified national framework, tertiary education in Germany is governed by the 16 federal states. Almost all universities are public institutions.\footnote{In 2024, more than 80 percent of all universities were public. See \citet{statistischesbundesamt2025} and \citet{statistischesbundesamt2026}.} Typically, professors at public universities are civil servants (``Beamte'').\footnote{There are cases in which professors are not civil servants but regular employees. The share in regular employment was 14 percent in 2019, compared to less than nine percent in the longer run \citep{detmer2019}.} This has implications with respect to the design of the academic labor market: First, there is a fixed number of positions that is determined by the government---with little to no scope for universities to hire someone unless a chair or professorship becomes vacant. Once tenured, professors are employed for life and benefit from very high employment protection legislation. That means that positions for new professors only open up when an incumbent retires or leaves for other reasons. Second, base salaries for civil servants are fixed by the federal state and vary only to a smaller extent by bonus payments for extraordinary achievements \citep[see e.g.,][]{preissler2025}. 

Receiving a call for a (tenured) professorship requires a long qualification period: Candidates graduate from university with a master's degree or equivalent. This is followed by doctoral studies which, during the time period of our study, were mostly organized at the chair and under the supervision of the professor/chair holder. In the absence of a centralized application procedure or graduate schools within departments, doctoral students were typically employed at the chair to work with the professor in research, teaching and administration. To qualify for a professorship, a successful doctoral dissertation would most often be followed by a ``habilitation'' which would again take several years.\footnote{Today, the process of habilitation has largely been replaced by ``habilitation-equivalent'' achievements such as longer postdoc periods or assistant professorships, either tenure-track or non-tenured. During the time period of our study, for somebody wishing to pursue an academic career, the expectation would have been to qualify through the habilitation process.} Once candidates qualified by obtaining the final (post-doctoral) degree of a ``habilitation'', they could apply for professorships when positions become vacant. There was no centralized application system and vacancies opened up at different points in time. The recruitment process of a professor in itself is lengthy in international comparison and can easily take more than one year.\footnote{The estimated average duration of the recruitment process from advertising until start lies between 14 months and two years \citep[see e.g.,][]{schade2025}.}

In summary, the German academic labor market is characterized by a number of key features: First, eligibility to become professor requires a long qualification period. This implies that adjustments in labor supply to labor demand shocks are slow, allowing us to isolate labor demand effects. Second, absent any targeted intervention, there is a fixed number of positions for professors (with fixed salaries), determined by the central government and not by the university. In other words, the academic labor market is strongly regulated. While this may differ from less regulated labor markets in the private sector, there are other public sector examples that are regulated in similar ways (e.g., the German labor market for teachers, judges or prosecutors). Third, professorships are considered to be attractive jobs, and labor supply typically exceeds demand leading to slack labor markets. Thus, a large number of new positions can actually be filled. Fourth, given decentralized application procedures, networks can be expected to be particularly important in job search and recruitment, leading to search frictions and, potentially, a non-diverse workforce. 

\subsection{The Natural Experiment: Historical Background}\label{subsec:history}

\begin{figure}[t]
    \centering 
    \begin{tabular}{c}
        \includegraphics[width=0.6\textwidth]{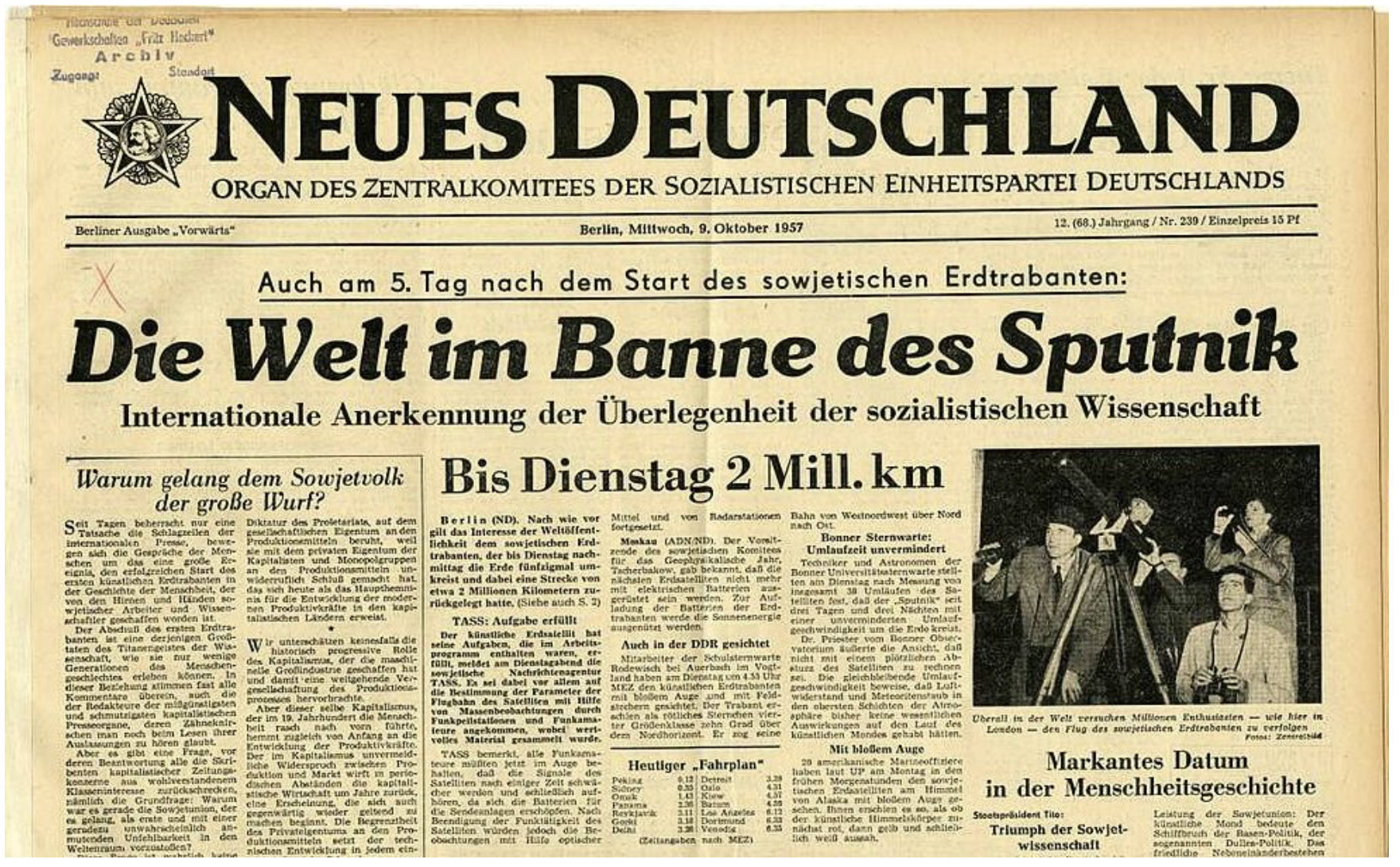}\\
    \end{tabular}
    \caption{
        \textbf{Sputnik shock and the ``superiority of socialist science'' in East Germany.} \\
        \footnotesize{
        \emph{Notes.}-  The figure presents a newspaper clip that covers the Sputnik shock in East Germany/\acf{GDR}. 
        \emph{Sources.}- Stiftung Haus der Geschichte, EB-Nr.\ OZS 001/1957/239. 
        }}
    \label{fig:history_news}
\end{figure}

The German reunification resulted in a natural experiment in the German academic labor market that allows us to empirically test the implications of a labor demand shock as postulated in our framework. During the 1945 to 1989 division of Germany, the \acf{FRG} in the West and the \acf{GDR} in the East had their own, distinct university systems. They did not only differ in terms of institutional design, but were, in fact, competing models at the time. Russia's launch of its first satellite---Sputnik---in 1957 illustrates the sentiment at the time. While the launch led to the ``Sputnik shock'' in the West, newspapers in East Germany ran headlines celebrating the ``superiority of socialist science'' (see Figure~\ref{fig:history_news}). 

From an organizational perspective, there were many differences regarding both research and teaching: The academic system in the \ac{FRG} was structured at the federal state level, and, following the ``Humboldtian Ideal'', based on the notion of unity of research and teaching \citep{guentherschmerbach}. In contrast, the \ac{GDR} had a centralized education system run by the state secretary. It was based on the Soviet educational system and the idea of a distinct separation between research and teaching. The latter was supposed to take place at institutes, while universities focused on educating students. Given the more student-centered and teaching-oriented focus, the \ac{GDR} saw comparatively higher staff numbers, especially on the pre-professorial level, to provide desired student-staff ratios \citep{kocka1994}. In terms of funding, professors in the East had fewer resources available for research than their counterparts in the West \citep{guentherschmerbach}. Important differences also existed in terms of academic freedom. While professors in the \ac{FRG} were given the constitutional right to freedom of research and teaching (Art.5 Absatz 3 GG), education and research in the \ac{GDR} were shaped by socialist ideology. For instance, there was a requirement for all doctoral students in the \ac{GDR} to submit proof that ``the knowledge of Marxism-Leninism acquired during their studies had been significantly deepened and expanded'' and appointments of personnel were subject to political stance and background.\footnote{\S 10(2) Anordnung zur Verleihung des akademischen Grades Doktor eines Wissenschaftszweiges - Promotionsordnung A - vom 21. Januar 1969. One famous example is Germany's former chancellor Angela Merkel, who - to fulfill these requirements - prepared a written work titled ``What is the socialist way of life?'', graded as ``sufficient'' (rite) by her ideology professor Joachim Rittershaus.}

Following the 1989 fall of the wall and the German Reunification in 1990, the East German higher education system was transformed. More specifically, the West German university model was transferred to the East.\footnote{A dual reform, combining the strengths of both systems, was dismissed as it would have required more time and financial resources, neither of which were considered available in the unforeseen situation \citep{guentherschmerbach}.} The Unification Treaty established the legal foundation for the transformation process which involved content, structural and personnel changes \citep{kehm1999}. In terms of content, the transformation aimed at the de-politicization of the East German research programs and teaching curricula. Structural changes meant that the university apparatus was downsized and disciplines were reorganized.

The main element of transformation, however, was the restructuring and replacement of academic staff \citep{pasternack1998}. The personnel changes started in 1990 with the closure of all departments of Marxism-Leninism, which had been introduced as a mandatory subject for any degree program in the \ac{GDR}. This was followed by the disbanding of politically affiliated institutions and the so-called ``Integrit{\"a}tsprüfungen'' (``integrity screenings''): Officially, all academic staff were screened and evaluated in terms of their scientific output, their political affiliation with the \ac{GDR} regime and the ideological content of the subject they taught. However, in practice the decision almost exclusively hinged upon the assessed ideological proximity of the subject to the socialist regime. Scientific personnel who failed the integrity screenings would not be retained on their previous position. Parallel to the dismissals, institutes and chairs were (re-)opened and staff was hired. This process was handled by founding and appointment committees (``Struktur- und Gründungskommissionen'', ``Berufungskommissionen'') who were in charge of finding qualified candidates.\footnote{Given the importance of the decisions, these committees were quite powerful and had a large impact on the university landscape in the former \ac{GDR}. Unfortunately, until today the exact working methods of the committees as well as the criteria for becoming a member remain opaque \citep{pasternack2021}. The committees consisted of representatives of professors (majority), researchers and students (Sächsisches Staatsministerium für Wissenschaft und Kunst: ``Erlass zur Umsetzung der Beschlüsse der sächsischen Staatsregierung''). The chairpersons of the founding committees (``Gründungsdekane'') were mostly males from the former West \citep{pasternack1998}. Decisions were taken by majority vote.}

\begin{figure}[t!]
    \centering 
    \begin{tabular}{cc}
        \multicolumn{1}{l}{\footnotesize{\textbf{Panel A.} Staff retention}} 
        & \multicolumn{1}{l}{\footnotesize{\textbf{Panel B.} Downsizing of staff}} \\
        \includegraphics[width=0.45\textwidth]{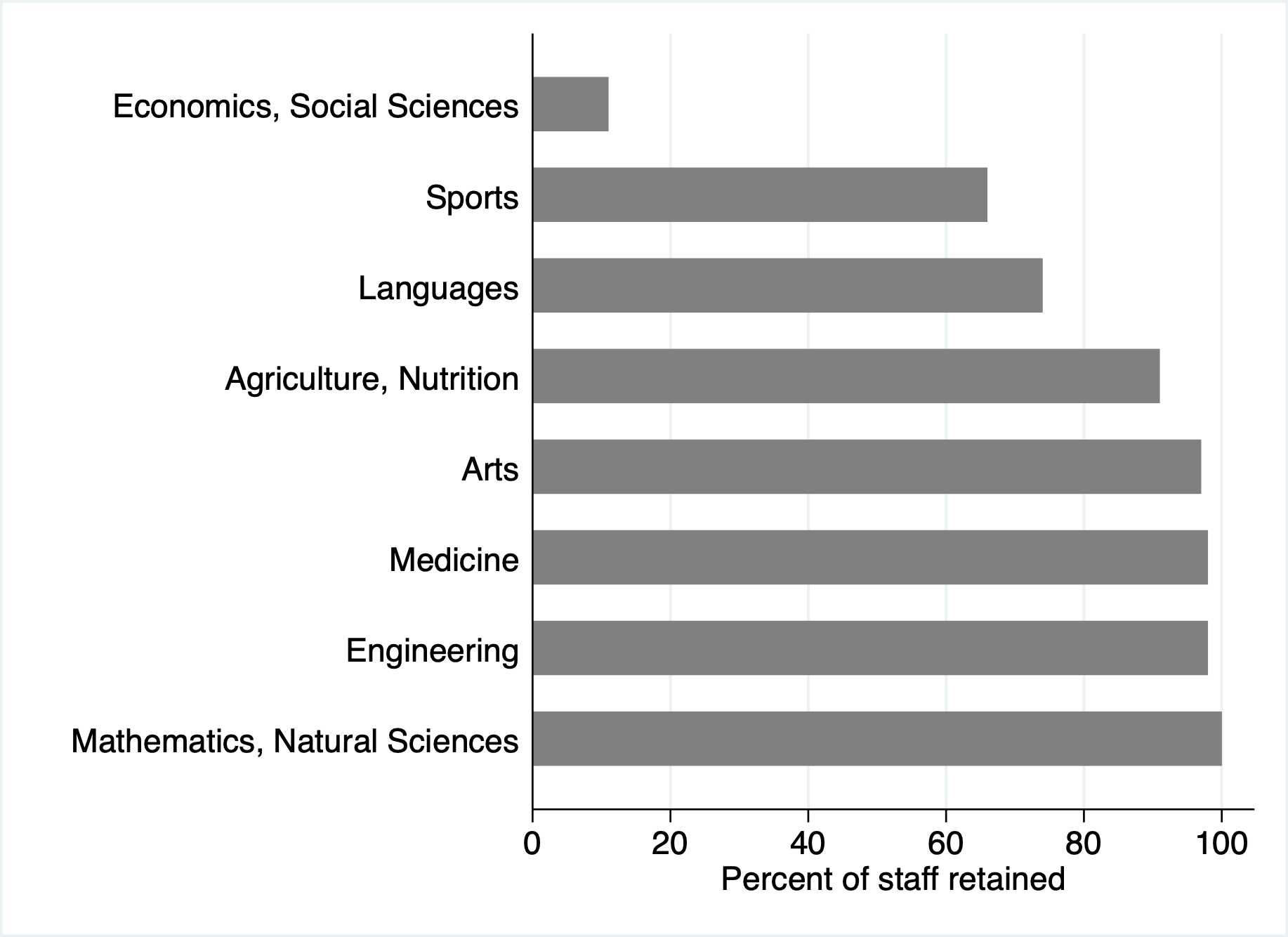} &
        \includegraphics[width=0.45\textwidth]{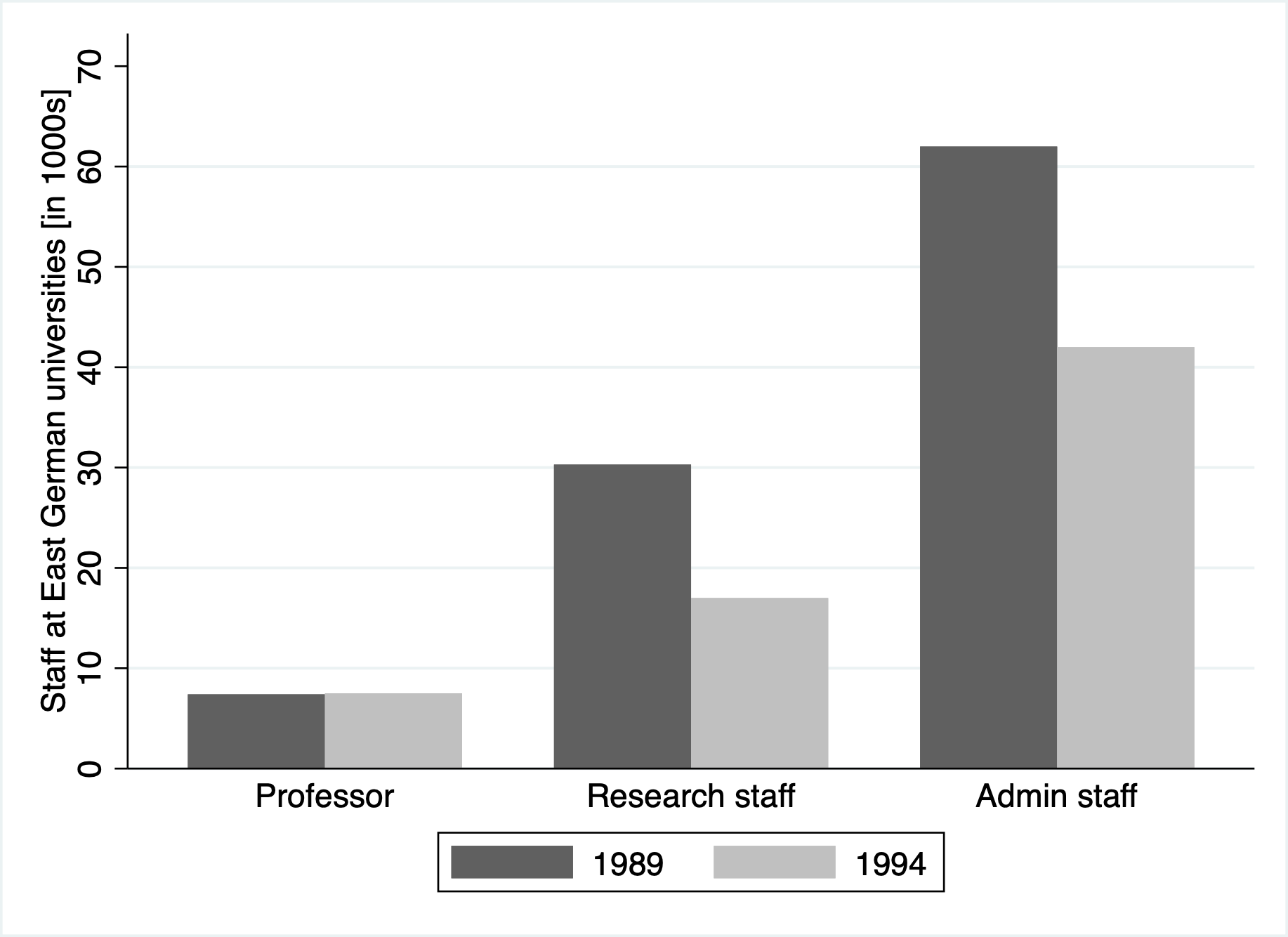}
    \end{tabular}    
    \caption{
        \textbf{Staff replacements at East German universities after the reunification.} \\
        \footnotesize{
        \emph{Notes.}- Panel A shows the percentage share of staff retained at universities in East Germany, by subject. Panel B shows the number of staff in East Germany (in thousands) by position (professors, research staff, administrative staff) before the transformation in 1989 and after the start of the transformation in 1994.
        \emph{Sources.}- Based on statistics from \citet{buckbechler1997} and own calculations.
    }}
    \label{fig:history_natexp}
\end{figure}

The extent to which different disciplines and subjects were affected by the transformation process varied substantially. Academics in subjects close(r) to politics and ideology, such as the social sciences, were affected most: Many scholars in the social sciences were negatively evaluated and dismissed, while those in the natural sciences or mathematics were perceived relatively ideology-neutral, facilitating a positive evaluation.\footnote{One example for these differences is a description of statistics in economics versus math departments \citep{strohe1996statistik}: \emph{``The strict control of statistics in terms of content only applied to economics, while mathematical statistics enjoyed extensive freedom in teaching and research as far as content was concerned. The appointment of personnel in both areas followed equally strict cadre development plans and similar requirements in terms of political stance and background. In the area of economics, the effects of this policy on the personnel structure were eliminated by the liquidations after 1990, with the exception of a few remnants, so that a new beginning was possible. In statistics at mathematical institutes, the old cadre policy may still have an effect for some time.''} Translated from ``Statistics in \ac{GDR} economic studies: between ideology and science'' (in German: Statistik im DDR-Wirtschaftsstudium: zwischen Ideologie und Wissenschaft), page 31.} Panel~A of Figure~\ref{fig:history_natexp} illustrates this: Only a small percentage of staff in economics and the social sciences were retained, compared to the vast majority in \ac{STEM} subjects. 
We interpret these differences between subject groups as differential labor market shocks and will exploit the resulting variation in our empirical analyses.

Besides the screenings, a significant downsizing of the East German university apparatus took place. Based on West German standards, the \ac{GDR}'s system was considered ``overstaffed''. Panel~B of Figure~\ref{fig:history_natexp} shows that the downsizing mainly affected research and administrative staff, while the number of professors remained rather stable in the aggregate. 
Starting directly after reunification, the personnel restructuring was concluded at the end of 1995 \citep{pasternack1998}. 

With downsizing and restructuring on the one hand and political and ideological screenings on the other hand \citep{kocka1994}, many professorships opened up. (Re-)opened positions were in principle open to scholars from the former West and the former East (or abroad). However, candidates from the former East often compared unfavorably, especially for leadership positions, as they struggled to meet the international standard evaluation criteria even if they passed the ``integrity screenings'' \citep{kehm1999}. We thus interpret the transformation process with many (re-)openings of professorships as an exogenous labor demand shock in East Germany, but mainly affecting candidates from West Germany. 

\section{Data and Research Design}

Our empirical analysis combines the implications derived from the conceptual framework with the variation that stems from the natural experiment described above.

\subsection{Data: Personnel Statistics for Academic Staff}\label{subsec:data}

We use data from the annual personnel statistics for academic staff by the German Federal Statistical Office \citep{destatis1998personal}. These are high-quality administrative microdata that contain information about the universe of non-administrative personnel at German universities. For all professors in a given year, we observe basic demographic information (gender, year of birth), information regarding their qualification (academic discipline, place of habilitation, year of first appointment as a professor) as well as their current employment (university, salary category). The data were entered by administrative personnel of the university, not by academics themselves, thus limiting selective non-response.\footnote{Further detail regarding the data can be found in \citet{janys2024}.} Our main analysis uses records from 1998 which is the first available year in the data. Although there is a three year gap between the conclusion of the personnel transformation process (1995) and the first time that we observe the new personnel composition, we do not consider this to be a serious constraint for our analysis given the very lengthy appointment process for professors in Germany. 

One limitation of our dataset is that it does not contain names; thus we cannot merge it to other individual-level data, such as citation databases. However, it nonetheless is the best choice to study German academic labor markets in the 1990s. This is due to two reasons: First, the data feature high-quality administrative data on all academic university personnel. It is not straightforward to create the universe of professors from other sources. Handbooks of academics, which can be used for earlier time periods, became increasingly self-selected as academics had to opt in to be featured. Departmental websites, on the other hand, had not been reliably published, especially for smaller universities (which, in contrast, are all included in our sample). Paper staff records at universities only need to be kept for a certain number of years, creating a new dimension of selection even if one was to go to every university and digitize every record still available.\footnote{Note that \citet{iaria2024} who provide the most complete picture of gender shares in academia in the 20th century stop their comprehensive analysis in 1969 and only add data for select disciplines at large universities for the year 2000.} 

Second, measuring quality is challenging -- especially when considering a large variety of subjects. Cross-subject comparisons are difficult when relying on citation databases for quality measures. These typically feature journal articles. Yet, different disciplines have different publication standards. While journal articles have been common in \ac{STEM} subjects for a long time, the social sciences have relied on book publications for much longer. These are generally not observable in publication databases (and, thus, nor are their authors). Even within fields that publish heavily in journals and are featured in these databases, the importance of a publication for an individual author as proxied by their position in the author list varies by subfield. Thus, citation databases cannot easily be used to generate the universe of academics nor to measure their quality across subjects. Instead, we propose a measure of perceived quality based on the place of the last postdoctoral qualification, which can be observed both at the time of hiring and today (see Section~\ref{subsec:academicmeasure} for details). 

\subsection{Evidence for Staff Replacements}\label{subsec:fs}

\begin{figure}[t!]
    \centering 
    \includegraphics[width=1.1\textwidth]{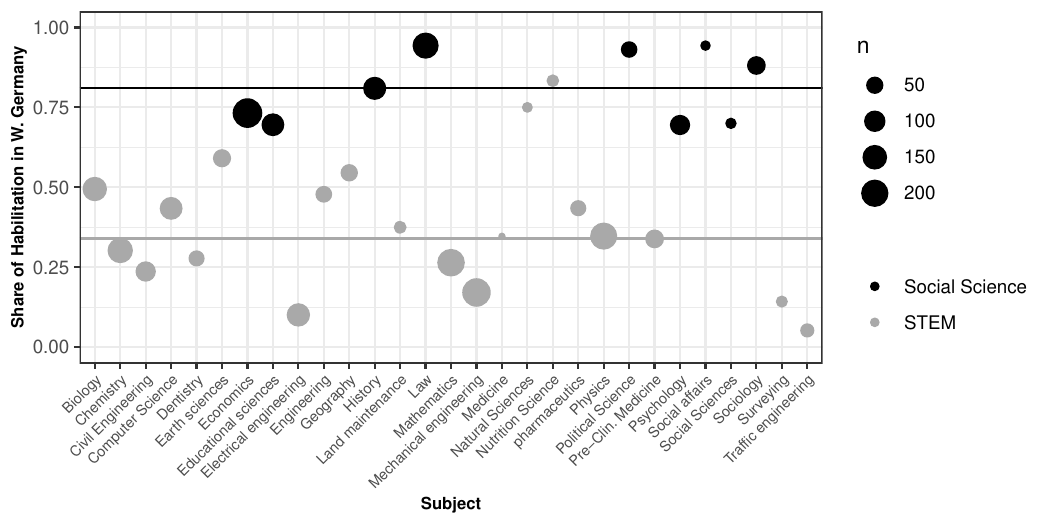}  
    \caption{
        \textbf{Professors in East Germany with a habilitation from West Germany.} \\
        \footnotesize{
        \emph{Notes.}- This figure shows the share of professors at East German universities in 1998 who hold a habilitation from a West German university, separately for each subject as indicated on the x-axis. Gray circles represent \ac{STEM}, black circles social sciences. The size of the circles is proportional to the size of the subject in terms of the professor count. The horizontal lines show the average share for \ac{STEM} (gray) and social Sciences (black), respectively. 
        \emph{Sources.}- Personnel statistics for academic staff \citep{destatis1998personal}, East German professors in 1998, and own calculations.
    }}
    \label{fig:fs1}
\end{figure}

Before turning to our research design and the core analyses, we look at our best approximation of a first stage in the data. As outlined above, we interpret the transformation process with the numerous (re-)openings of academic jobs at East German universities as an exogenous labor demand shock mainly for those candidates with a habilitation from West Germany. This is, in fact, something that we can observe in the data: If the replacement positions were mainly filled by West Germans, we would see a larger share of West Germans in the social sciences compared to \ac{STEM} fields in 1998.

Figure~\ref{fig:fs1} plots the share of professors at East German universities in 1998 who had obtained their habilitation from a West German university. Black circles represent the shares for social sciences subjects, gray ones for \ac{STEM} subjects. The size of the circles corresponds to the size of the different subjects in terms of the number of professors. Starting with \ac{STEM}, around one third of those who were professor at an East German university in 1998 had received their habilitation in the West, with some variation across individual subjects. With about 80 percent, this share is much higher in the social sciences. We interpret these observations as a quasi-first stage: The social sciences (treatment group) have a substantially higher share of professors with West German qualifications than \ac{STEM} (control group), and this looks quite homogeneous across specific subjects within the social sciences. These findings are consistent with the notion that when new positions opened up due to the staff replacements, they were filled with aspiring candidates moving from the West to the East.\footnote{Figure~\ref{fig:fs1} shows the share of professors at East German universities who had obtained their habilitation from a West German university. The opposite share of professors at West German universities with an East German habilitation is very low (in 1998) at 0.02. That is, while there is some mobility of scholars from the former \ac{GDR} to the West, this is extremely rare.}\textsuperscript{,}\footnote{A concern could be that the social sciences and \ac{STEM}, respectively, had a different subject composition in the East and in the West. When looking at the share of each subject in the respective field, we find that in 1998 the relative importance of the subjects is similar across the East and the West, with a correlation of 0.9.}

\begin{figure}[t!]
    \centering 
    \includegraphics[width=0.6\textwidth]{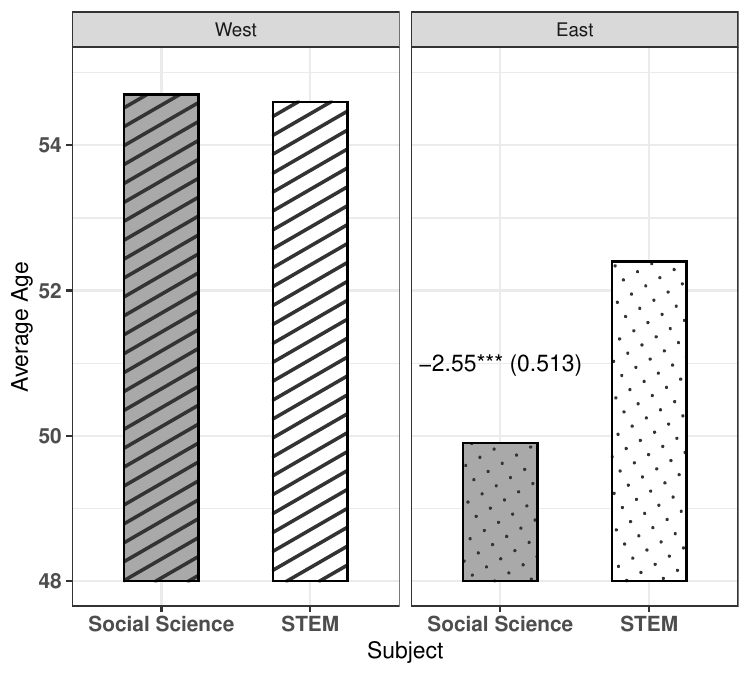}   
    \caption{
        \textbf{Average age of professors across subjects in East and West Germany.} \\
        \footnotesize{
        \emph{Notes.}- This figure plots the average age of professors for West Germany (left, lines) and East Germany (right, dots), and for social sciences (gray) and \ac{STEM} (white), respectively. The coefficient displayed is the estimate of $\delta$ in equation~(\ref{eq:exp2_did2}) with age as the dependent variable. The associated standard error, clustered by subject, is displayed in parentheses. 
        \emph{Sources.}- Personnel statistics for academic staff \citep{destatis1998personal}, professors in 1998, and own calculations.
    }}
    \label{fig:fs2}
\end{figure}

Figure~\ref{fig:fs2} plots the raw differences in the average age of professors who, in 1998, worked at West German (left) or East German universities (right), respectively.\footnote{Similar figures are available for subsequent years 1999-2002 but not shown in the paper.} Gray bars show averages for social sciences and white bars for \ac{STEM}. Professors at West German universities were on average just under 55 years old in 1998. There is no noticeable difference between social sciences and \ac{STEM}. Looking at East German universities, two observations stand out. First, professors in both groups are on average younger than in the West. Second, professors who worked in the social sciences at East German universities in 1998 are on average younger than those in \ac{STEM}: The average age in \ac{STEM} is about 52.5 years, compared to just below 50 years in social sciences. This difference between groups in the East---in comparison to the lack of a difference between groups in the West---suggests that on average younger candidates were hired during the staff replacement process at East German universities. The fact that the positive labor demand shock, induced by the staff replacements, led to a decrease in the average age is informative: It signals that indeed there was a ``first stage'': as positions opened up, new (younger) professors were hired from the West German pipeline.

\subsection{Research Design: Two-Way Fixed Effects}\label{subsec:academicdesign}

Our empirical approach leverages the differences in staff replacement rates across subjects. As shown above, while only about ten percent of professors in economics and the social sciences were retained in East Germany after the transformation, the vast majority of professors in \ac{STEM} subjects kept their positions. We explore these stark differences in the intensity of the transformation and replacement processes in a two-way fixed effects approach (spatial \acl{did} design). We define social sciences as the treatment group and \ac{STEM} as the control group.\footnote{Social Sciences include: Economics, Educational Sciences, History, Law, Political Science, Psychology, Social Affairs, Social Sciences. \ac{STEM} include: Biology, Chemistry, Civil Engineering, Computer Science, Dentistry, Earth Science, Electrical Engineering, Geography, Land Maintenance, Mathematics, Mechanical Engineering, Medicine, Natural Sciences, Nutrition Science, Pharmaceutics, Physics, Pre-Clinical Medicine, Surveying, Traffic Engineering.} Instead of a time dimension, as in most common \acl{did} settings, we explore the spatial dimension and compare West Germany (analogue to ``pre'') and East Germany (analogue to ``post''). This allows us to net out baseline differences between the social sciences and \ac{STEM}. The most basic \acl{did} design can then be represented as follows:

\begin{equation}
    Y_{i} = 
        \alpha + \gamma \cdot \texttt{SocSci}_{i}+\lambda \cdot \texttt{East}_{i} + \delta \cdot (\texttt{SocSci}_i \times \texttt{East}_i)+u_{i}
    \label{eq:1}  
\end{equation}

where $\texttt{SocSci}$ equals one if individual $i$ is a professor in the social sciences and zero if they are a professor in \ac{STEM}. $\texttt{East}_i$ equals one if they are a professor in East Germany and zero if in West Germany. $Y_i$ denotes the outcome which measures different characteristics---academic background and gender. Under the conventional assumption of linear additivity plus more critical assumptions regarding treatment exogeneity and the unobserved counterfactual (discussed below), we can identify the causal parameter $\delta$: 
\begin{align}
    \delta &&= 
         \left[ \mathbb{E}[Y_i|\texttt{SocSci}=1,\texttt{East}=1] - \mathbb{E}[Y_i|\texttt{SocSci}=1,\texttt{East}=0] \right] \nonumber \\
        && - \left[ \mathbb{E}[Y_i|\texttt{SocSci}=0,\texttt{East}=1] - \mathbb{E}[Y_i|\texttt{SocSci}=0,\texttt{East}=0] \right]
    \label{eq:exp2_did2}
\end{align}

with the sample analogue 
$\hat{\delta} = 
        \left( \bar{Y}_{\texttt{SocSci}}^{\texttt{East}} - \bar{Y}_{\texttt{SocSci}}^{\texttt{West}} \right) - 
        \left( \bar{Y}_{\texttt{STEM}}^{\texttt{East}} - \bar{Y}_{\texttt{STEM}}^{\texttt{West}} \right) 
$.

In practice, we estimate regressions based on equation \eqref{eq:1} but instead of a binary variable for the social sciences versus \ac{STEM}, we include subject $s$ fixed effects. This allows for more fine-grained time-invariant differences across subjects, e.g.,\ differences in the female share between physics and biology. For ease of notation, we also denote the binary variable indicating East versus West Germany as a region $r$ fixed effect. Our main regression equation can thus be written as:
\begin{align}\label{eq:exp2_did1}
    		Y_{i} = \gamma_{s}\mathbb{I}\{i \in S=s\} + \lambda_r\mathbb{I}\{i \in r=\texttt{East}\} + \delta \cdot \mathbb{I}\{i \in \texttt{SocSci}\} \times \mathbb{I}\{i \in \texttt{East}\}+u_{i}
	\end{align}
where $i: 1,...,N$ indexes the individual, $s:1,...,S$ is the subject, $r: \texttt{East, West}$ is the region. $\mathbb{I}\{i \in \texttt{SocSci}\}$ equals one if individual $i$ is a professor in the social sciences and zero if they are a professor in \ac{STEM}. Standard errors are clustered at the subject level. The outcome $Y: \texttt{academic background, gender}$ measures the two dimensions of our framework: demographic type and perceived quality. More precisely, in the case of demographics, we define $Y$ as a binary variable equal to one if the individual is female, and equal to zero otherwise. In the case of perceived quality, we do not directly observe the quality of a professor. Instead, we proxy it by the habilitation department of the individual. Because there was no clear and well-published ranking of German universities (or departments) at the time, we cannot use the rank of the habilitation department as the direct outcome. Instead, we leverage the dispersion (versus concentration) of habilitation departments as a proxy for subject-level (perceived) quality. Details are explained in Section~\ref{subsec:academicmeasure}, but the intuition is as follows: A high concentration in habilitation departments indicates that professors are mainly hired from select places which are perceived as high quality. Similarly, an increase in the dispersion of habilitation departments implies that professors are also hired from lower-ranking places. We implement this notion by constructing an index that quantifies the dispersion/concentration within subjects (see Section~\ref{subsec:academicmeasure}). We estimate equation~(\ref{eq:exp2_did1}) at the individual level.\footnote{Information on habilitation department is not available for all individuals in the sample. More precisely, for a share of professors the habilitation department is coded as ``other German University''. We drop those from the regressions with our perceived quality outcome. This explains why the number of observations in Panel~A of Table~\ref{tab:did} is smaller than in Panel~B for the gender outcome. 
}

To interpret the coefficients causally, we need to assume that, in the post-reunification era, any East-West differences in the outcome gaps between social sciences and \ac{STEM} are only caused by differential replacement rates for professors in the aftermath of the reunification. In other words, we need to assume that West Germany proxies the (unobserved) counterfactual staff composition in East Germany without replacements well. This implies two specific assumptions. 

First,  we need to assume that differences in quality and gender composition between the social sciences and \ac{STEM} would have been the same in East and West Germany had the replacements in the East not taken place. This is similar to the parallel trends assumption in the temporal \ac{did}-framework, i.e., we assume that the untreated potential outcomes are independent of receiving the treatment. Because of the partially spatial setting, unlike in a temporal \ac{did} framework, we cannot provide empirical evidence in the form of ``pre-trends'' for this. However, at least for the dimension of gender, we can provide some empirical evidence on what this ``alternative universe'' (without the re-unification) would have looked like.\footnote{Unfortunately, for the quality dimension this is much harder due to data restrictions.} The intuitive way to do this would be to look at the pre-reunification period; however, there is no high-quality data before 1990 to reliably compare gaps in the social sciences versus \ac{STEM} share of female professors between East and West Germany. But, anecdotally we know that even though enrollment of women in \ac{STEM} subjects was higher in East than in West Germany, there was still---as in West Germany---a large and persistent gap (by subject) between male and female students in overall enrollment and in PhD completion. For example, between 1975 and 1980, around half of the doctoral theses in literature, linguistics and medicine in the \ac{GDR} were written by women, compared to only 5.2\% in the technical sciences \citep{budde2003frauen}.
Looking at the aggregate numbers of professors prior to reunification, the gap in the share of female professors between social sciences and \ac{STEM} was higher in the East (9.2~percentage points difference) than in the West (5.4~percentage points difference).\footnote{These numbers are taken from \citet[][Table~3.4]{statistischesbundesamt1990} for the West and from \citet[][Table~2.2]{burkhardt1997} for the East. West German numbers refer to full-time professors in C-payment schemes in 1989 at all higher education institutions. East German numbers refer to full-time professors in 1989 at all higher education institutions. Note that aggregate numbers make it hard to make East and West German professors completely comparable as there were different institutional settings according to which aggregates were formed (e.g., by type of higher education institution) and different sub-disciplines were differently popular in the East and the West.} In other words, in the West, the share of female professors was 5.4 percentage points higher in the social sciences than in \ac{STEM} subjects; in the East, 9.2 percentage points. Thus, had the reunification not taken place, we would have expected the gap in the share of female professors between the social sciences and \ac{STEM} subjects to be around 3.8 percentage points (9.2 minus 5.4) larger in the East than the West in 1998. In that sense, estimates of $\delta$ in equation~(\ref{eq:exp2_did1}) that are smaller (larger) than 0.038 imply that the social sciences in East Germany became more (less) similar to West Germany, equivalent to a reduction (increase) in gender diversity in the social sciences compared to the pre-reunification period.

The above argument only holds if outcomes would not have changed in the East German social sciences regardless of reunification and the subsequent staff replacements. We argue that this is plausible considering the institutional context of the \ac{GDR}: Young researchers could not choose fields and women were under-represented both in professorship positions and in the academic pipeline. Specifically, the \ac{GDR} employed a centralized system in which students were assigned to subjects according to the projected labor demand and not based on personal interests. A similar system decided who was allowed to continue with an academic career. Although women were expected to work outside the home and there were no formal barriers to their education in the \ac{GDR}, women remained largely under-represented in leadership roles, academia being no exception despite formal equality \citep{ross1999verhinderter}. Additionally, the \ac{GDR} also had a ``leaky pipeline'': In 1986, the share of women who graduated with a PhD was 31.8\%; 
the share of habilitations written by women only 14.9 \% (see Table~\ref{tab:diss_women_gdr}). Given this context, it is unlikely that the difference in the gender gap between \ac{STEM} and the social sciences in the East would have further narrowed independent of the staff replacements.

Second, analogously to the \acf{SUTVA}, we need to assume that there are no spillovers. This is particularly important as we exploit spatial variation. Specifically, we need to assume that outcomes in \ac{STEM} in the East did not change because of the staff replacements. Given the high retention rates documented in the historical records, this assumption seems plausible. Similarly, we need to assume that outcomes in the social sciences in the West did not change because of the staff replacements in the East. This is a more critical assumption, as it is in theory possible that incumbent (male) professors from West Germany moved to the East when positions opened up there and that their initial positions were subsequently filled with (younger) female or lower-quality professors. This could have led to an (implicit) replacement wave in the West with an increase in diversity in either or both dimensions. As a result, our estimates might be attenuated. We note, however, that Figure~\ref{fig:fs2} already showed a significant drop in the average age of the professors in the new positions in the East---which suggests that younger rather than older, more experienced candidates from West Germany were hired. 

If the above assumptions hold, $\delta$ in equation~(\ref{eq:exp2_did1}) captures the effect of a labor demand shock without simultaneous changes in labor supply. As the profession of professors requires a long qualification period that includes the completion of a habilitation, it is unlikely that workers from the private sector could easily react to a labor demand shock in academia. Our setting is thus well suited to study the quality and gender diversity implications of a local labor demand shock.


\section{Compositional Changes in (Perceived) Worker Quality}\label{sec:workerchars}

We use the natural experiment and research design described above to take the implications of our framework in Section~\ref{sec:model} to the data. In this section, we start by looking at the worker characteristics $C$, the quality signal. Recall that our framework implied that in a scenario with a ``filled pipeline'', the worker composition would be kept constant in both dimensions (quality and gender). But, in a scenario where this is not possible, there was a trade-off between the demographic and the quality dimension. In this case, the model predicts that the share of high-quality type workers decreases with a demand shock if the status-quo demographic composition is held constant. In this section, our aim is thus to test whether the post-demand shock quality composition of the workforce differs from the status quo. That is, for workers of quality type $c\in C$, we test whether

$$ q_{\cdot c}^{\Delta} = \tilde{q}_{\cdot c} $$

We measure $C$ using the academic background of professors. The academic background is a relevant characteristic in this context where professional networks and the reputation of the home institution are likely to be interpreted as quality signals by employers: If selection at the home institution was, at least to some degree, meritocratic, the average quality (reputation) of the home institution signals applicants' individual quality. In addition, departments with a higher reputation might also directly affect academic production \citep[see e.g.,][]{bethmann2023, waldinger2010}---for example, through skills taught, high-quality networking opportunities and resulting cooperation and publication opportunities---, making an applicants' academic network part of their quality characteristics. As such, academic background can plausibly be interpreted as signal of worker quality in our context. Empirical patterns suggest that hiring committees indeed use academic background as a quality signal. For example, almost 60\% of faculty at the top 96 US economics departments received their PhD from a top-15 university \citep{jones2024}. We proxy the academic background of professors by their university of habilitation, i.e., the university from which they obtained their final (postdoctoral) qualification to become professor.

\subsection{Measurement of Quality: Dispersion of Academic Backgrounds}\label{subsec:academicmeasure}
 
The first step is to create a quantitative measure of (perceived) quality. As highlighted above, this is difficult as there was no clear and well-published ranking of universities, let alone departments, for this time period in Germany. Instead, we use the dispersion (or concentration) of professors' habilitation departments. A high concentration in habilitation departments indicates that professors are mainly hired from select places which are perceived as high quality departments. Similarly, an increase in the dispersion of habilitation departments implies that professors are also hired from lower-ranking places. 

The key challenge to quantify the dispersion in our context is that the total number of universities from which individuals could have received their habilitation is unknown. This is partly because departments may change over time, but also because scholars may have obtained their qualifications abroad. Statistically, this means that we cannot apply commonly used concentration indices, such as the \acf{HHI}, as we do not observe the total number of types.\footnote{An additional concern could be non-randomly missing information regarding the habilitation, maybe because norms or requirements for this qualification changed over time. In the data, the correlation between missing habilitation information and age is 0.66, i.e.,\ older cohorts are more likely to not have a recorded habilitation. This is in line with habilitation standards being more stringently implemented in later years, including our sample period. When looking at correlations of missing academic age (i.e.,\ year of habilitation) within age of birth group and across East and West Germany and comparing social sciences and \ac{STEM}, we find that missing habilitation data is mainly age and subject driven, but unrelated to East and West dimensions.} 

Instead, we develop a new index accounting for this. We define our dispersion index $D$ for subject $s$ in region $r$ (East/West) as:
\begin{align}
    D^{r}_{s}=\frac{(le(x_s)-1)}{N_s+\sigma_x-1},
    \label{eq:index}
\end{align}
where $x_s$ is a vector of the distribution of professors in subject $s$ across habilitation universities. For example, if there were four economics professors of whom one got their habilitation in Mannheim and three in Bonn, the vector would read $x_s=(1 \enspace 3)$. The dispersion measure $D$ increases in $le(x_s)$ which is the length of the vector and represents the number of distinct universities from which professors in subject $s$ received their habilitation. As $le(x_s)$ increases with the number of professors in subject $s$ (without measuring additional diversity), we scale the index by taking into account $N_s$, the total number of professors in subject $s$. In addition, we scale $D$ by $\sigma_x$, the standard deviation of $x_s$, to account for the fact that professors from the same number of habilitation departments can be more (or less) equally distributed across these places, indicating more (less) dispersion in academic background. Thus, $D_s$ decreases in $N_s$ and in $\sigma_x$. 

\begin{figure}[t!]
    \hspace{-1cm}
    \centering    
    \begin{tabular}{cc}
        \multicolumn{1}{l}{\footnotesize{\textbf{Panel A.} Minimum dispersion}} 
        & \multicolumn{1}{l}{\footnotesize{\textbf{Panel B.} Maximum dispersion}}\\   
        \includegraphics[width=0.45\textwidth]{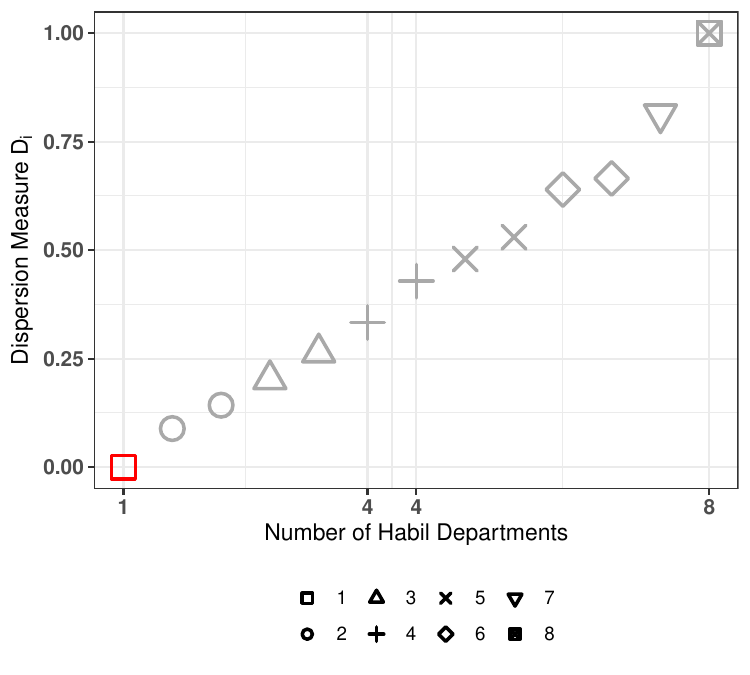} 
        & \includegraphics[width=0.45\textwidth]{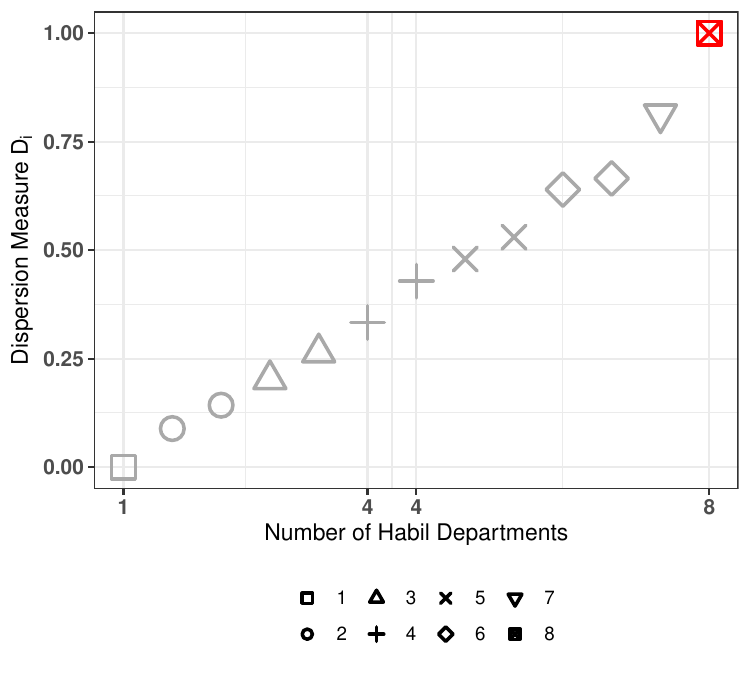}\\ 
        \multicolumn{1}{c}{\footnotesize{\textbf{Panel C.} Some dispersion}}\\
        \includegraphics[width=0.45\textwidth]{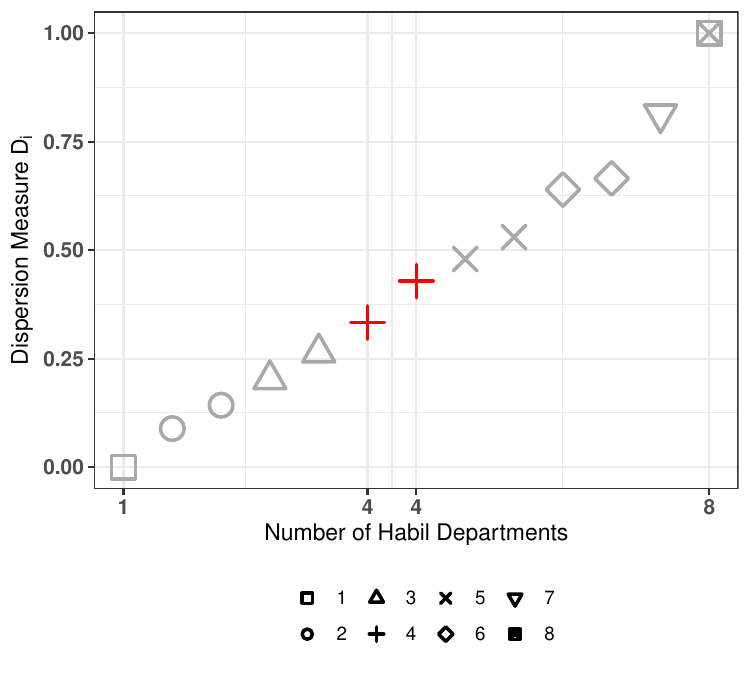} \\
    \end{tabular}    
    \caption{
        \textbf{Dispersion index examples.} \\
        \footnotesize{
        \emph{Notes.}- The figure illustrates examples of our dispersion measure from equation~(\ref{eq:index}) for a case with eight professors, distributed sequentially to up to eight universities from where they obtain their qualification (i.e., habilitation). Panel~A shows the example of minimum dispersion (qualification from the same university, dispersion index of 0 indicated by the red box in the bottom left corner). Panel~B shows the example of maximum dispersion (qualification each from another university, dispersion index of 1 indicated by the red crossed box in the top right corner). Panel~C shows two examples with some but not full dispersion and different levels of concentration (qualification from different universities, dispersion index of 0.333 and 0.429 as indicated by the plus red crossed in the middle). See the text for more detailed explanations.
        \emph{Sources.}- Based on own calculations.
    }}
    \label{fig:index}
\end{figure}

The qualitative interpretation of our dispersion measure is straightforward: The higher $D_s$, the higher is the dispersion of academic backgrounds (and hence of perceived quality). The quantitative interpretation is less obvious and best illustrated through examples. Let us consider a subject $s$ with eight professors in total. In the first case, illustrated in Panel~A of Figure~\ref{fig:index}, all eight professors received their habilitation from the same university: $x_s=(8)$ and $le(x_s)=1$, $N_s=8$, $\sigma_x=0$ such that $D_s=0$. There is maximum concentration of the academic background in this subject and the index takes the minimum value of zero (bottom left corner in the figure). In the second case, illustrated in Panel~B, we take the opposite extreme and imagine a case in which each of the eight professors received their habilitation from a different university: $x_s = (1\enspace 1\enspace 1\enspace 1\enspace 1\enspace 1\enspace 1\enspace 1)$ and $le(x_i)=8$, $N_s=8$, $\sigma_x=0$ such that $D_s=1$. There is full dispersion of the academic background in this subject and the index takes the maximum value of one (top right corner in the figure). Finally, we assess less extreme cases to illustrate how our dispersion measure works. Again, we consider the subject $s$ with eight professors, this time with four habilitation universities. In one case, five professors received their habilitation from the same university and the other three from three additional universities: $x_s=(5\enspace 1\enspace 1\enspace 1)$ and $le(x_s)=4$, $N_s=8$, $\sigma_x=2$ such that $D_s=0.333$. In the other case, two professors each received their habilitation from the same university: $x_s=(2\enspace 2\enspace 2\enspace 2)$ and $le(x_s)=4$, $N_s=8$, $\sigma_x=0$ such that $D_s=0.429$. That is, dispersion according to our index is higher in the second case---taking into account not only how many habilitation universities are represented but also the concentration across these universities. Finally, we note that our measure is scale dependent: If the total number of professors $N_s$ exceeds the possible number of habilitation universities, then the dispersion index cannot reach the maximum value of one and $D_s<1$ even if all possible habilitation universities were represented. While this complicates the interpretation of our results in absolute terms, we can harmonize differences across subjects and regions by including subject and region fixed effects in our regressions.

\subsection{Results: Changes in Dispersion of (Perceived) Quality}\label{subsec:academicresults}

\begin{figure}[t!]
    \centering 
    \begin{tabular}{cc}
        \multicolumn{1}{l}{\footnotesize{\textbf{Panel A.} Dispersion index}} &
        \multicolumn{1}{l}{\footnotesize{\textbf{Panel B.} Gender}}\\
        \includegraphics[width=0.48\textwidth]{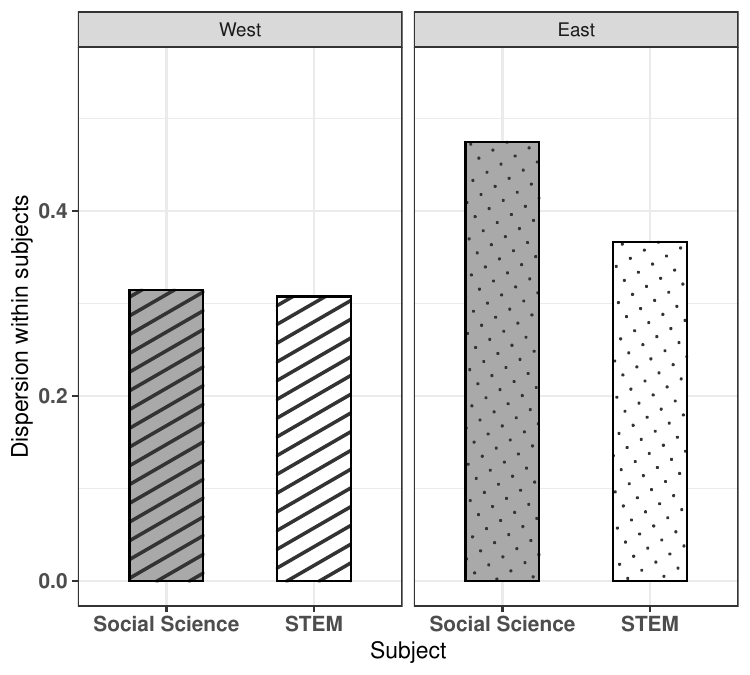}
        &
        \includegraphics[width=0.48\textwidth]{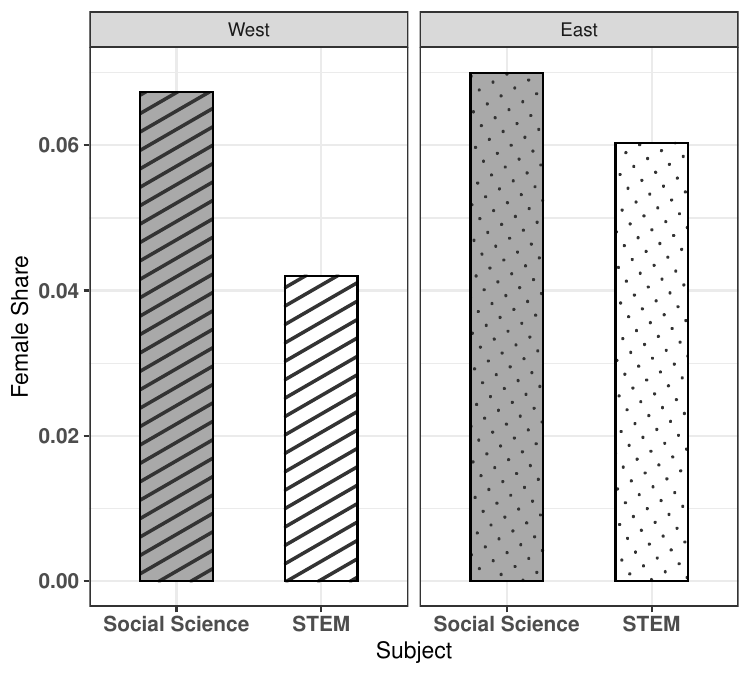}   
    \end{tabular}    
    \caption{
        \textbf{Raw differences in outcomes.} \\
        \footnotesize{
        \emph{Note.} - The figures show the average of the dispersion index (Panel~A) and the share of female professors (Panel~B) for West Germany (left, lines) and East Germany (right, dots), as well as for \ac{STEM} subjects (gray) and social sciences subjects (white), respectively. 
        \emph{Sources.}- Personnel statistics for academic staff \citep{destatis1998personal}, professors in 1998, and own calculations.
    }}
    \label{fig:outc_raw}
\end{figure}

Panel~A of Figure~\ref{fig:outc_raw} shows the raw differences in dispersion of academic backgrounds for professors in West Germany (left) and East Germany (right) in 1998. The gray bars indicate the dispersion in social sciences, the white bars in \ac{STEM} subjects, respectively. In West Germany, the dispersion index lies at just over 0.3 for both subject groups. In East Germany, dispersion is higher in \ac{STEM} than in the West (0.367 vs. 0.308) but again observably higher in social sciences (0.475 vs. 0.315). That is, the differences in the raw data suggest higher dispersion in the treatment group (social sciences) relative to the control group (\ac{STEM}) with a raw gap of 0.101 index points.

Given the almost equal dispersion in academic backgrounds in social sciences and \ac{STEM} in West Germany, one would expect the regression results corresponding to equation~(\ref{eq:exp2_did1}) to be close to this raw gap. Indeed, this is what we find. The results for the regressions are shown in Panel~A of Table~\ref{tab:did} where column~(1) corresponds to the baseline specification. We find that dispersion in the social sciences is about 0.12 index points higher than in \ac{STEM} in East Germany net of the (in this case very small) baseline differences between the subjects measured in West Germany. The estimate is statistically significant at the 1~percent level. 

A comparison to the averages shown in Panel~A of Figure~\ref{fig:outc_raw} already suggests that a difference of 0.12 is meaningful (33 percent relative to the \ac{STEM} mean in the East). A simple simulation exercise further puts the magnitude into context. For this exercise, we consider economics professors in West Germany.\footnote{Economics is broadly defined here according to the German definition ``Wirtschaftswissenschaften'' which includes Economics but also Management/Business.} In 1998, there were 602 economics professors in West Germany. The top ten habilitation universities in economics were: Cologne (50), Mannheim (36), Muenster (33), Munich (29), Hamburg (26), Goettingen (24), Karlsruhe (24), Bonn (23), Frankfurt (22) and Free University Berlin (22). The calculated dispersion index is $D_{s=\texttt{Economics}}^{r=\texttt{West}}=0.09$. Keeping the total number of professors constant, we simulate a scenario in which the number of habilitation universities is doubled. This results in a simulated dispersion index of $D_{s=\texttt{Simulated Economics}}^{r=\texttt{West}}=0.183$. In other words, an increase in the dispersion measure by 0.12 index points corresponds to a slightly larger change than from doubling the number of distinct habilitation universities in economics (0.183 - 0.09 = 0.093).
    
\begin{table}[t!] 
     \centering
     {\small
\begin{tabularx}{\textwidth}{l *{3}{>{\centering\arraybackslash}X}}
\toprule
&{(1)}&{(2)}&{(3)} 
\tabularnewline
\addlinespace[\belowrulesep]
\textbf{Panel A. Perceived quality} &  \textbf{Dispersion index} & \textbf{Inverse normalized HHI } & \textbf{Triple interaction} \tabularnewline \cmidrule{2-4}
& \emph{Baseline} & \multicolumn{2}{c}{\emph{Additional specifications}} \tabularnewline
\addlinespace[\belowrulesep]
$\texttt{SocSci}_i \times \texttt{East}_i$  &        0.1196$^{***}$   &0.0860$^{***}$&  0.1257$^{***}$  \tabularnewline
                                            &    (0.0393)      &  (0.0000)         &  (0.0397) \tabularnewline
$\texttt{SocSci}_i \times \texttt{East}_i \times \texttt{Female}$  &&&-0.0334\tabularnewline
           &&&                         
                               (0.1310)            
\tabularnewline \cmidrule{2-4}                                         
N & 7,866 & 7,866 & 7,866\tabularnewline
N with non-German habilitation & 229 & 229 & 229 \tabularnewline
\midrule 
\addlinespace[\belowrulesep]
&{(4)}&{(5)}&{(6)} \\
\textbf{Panel B. Y: Gender} & \textbf{Baseline} & \textbf{Cohort FE} & \textbf{Academic Age FE}\tabularnewline \cmidrule{2-4}
& \emph{Baseline} & \multicolumn{2}{c}{\emph{Additional specifications}} \tabularnewline
\addlinespace[\belowrulesep]
$\texttt{SocSci}_i \times \texttt{East}_i$  &-0.0122     & -0.0057   &-0.02145\tabularnewline
                                            &   (0.0113)      & (0.0136)          &  (0.0161)   \tabularnewline
\tabularnewline \cmidrule{2-4}                                       
N & 16,589 & 16,589& 16,589 \tabularnewline
\bottomrule 
\end{tabularx}
}
     \caption{
        \textbf{Results from the two-way fixed effects estimation.}\\
        \footnotesize{
        \emph{Notes.}- This table presents coefficient estimates of $\delta$ in equation~(\ref{eq:exp2_did1}). Panel~A shows the results when the outcome is the dispersion of academic backgrounds. Columns~(1) and (3) are based on our dispersion index from equation~(\ref{eq:index}), column~(2) on an inverse normalized \acf{HHI}. Columns~(1) and (2) refer to our baseline specification, column~(3) additionally interacts the social sciences-East interaction term with a binary variable indicating whether the individual is female. Panel~B shows the results for the gender composition when the outcome is a binary variable indicating whether the individual is female. Column~(4) shows the baseline specification, column~(5) adds year of birth (cohort) fixed effects and column~(6) year of habilitation (academic age) fixed effects. Standard errors are clustered by subject and displayed in parentheses. *** p$<$0.01, ** p$<$0.05, * p$<$0.1.
        \emph{Sources.}- Personnel statistics for academic staff \citep{destatis1998personal}, professors in 1998, and own calculations.
        }
    }
    \label{tab:did}
\end{table}

\paragraph{Robustness.} We conduct two robustness checks for the results presented here. First, one might be concerned that the estimates are driven by specific subjects within social sciences and/or \ac{STEM}. Appendix Figure~\ref{fig:disp_subj} suggests that this is not the case: It illustrates the variation in the dispersion index by subject and highlights that the higher dispersion in social sciences compared to \ac{STEM} is not driven by singular disciplines. Second, instead of our dispersion measure, we use an inverse normalized \acf{HHI}. The results are shown in column~(2) of Panel~A in Table~\ref{tab:did}. They confirm our previous conclusions. 

\paragraph{Interpretation.} Taken together, the results suggest a significant and robust increase in dispersion with respect to academic backgrounds as a result of the labor demand shock. This indicates that the academic labor market opened up to candidates from lower-ranking universities and the new positions in the East were (also) filled by candidates with lower quality signals. In terms of the framework, this means that we can reject that $q_{\cdot c}^{\Delta} = \tilde{q}_{\cdot c}$, and instead we find evidence that is consistent with $q_{\cdot H}^{\Delta} < \tilde{q}_{\cdot H}$, a decrease in the share of workers of the (perceived) high-quality type. 


\section{Compositional Changes in Worker Demographics}\label{sec:workerdem}

Next, we turn to the second dimension in our model, the demographic characteristic $D$. Again, our framework in Section~\ref{sec:model} implied that the worker composition would not change in response to the demand shock if the pipeline was filled. Section~\ref{sec:workerchars} considered the scenario in which this is not possible and the resulting trade-off is solved by substituting to lower-quality types while keeping the demographic composition constant. In this section, we consider the alternative scenario in which potential losses in average worker quality are mitigated by substituting away from the predominant demographic type as a way to mitigate potential quality losses associated with a demand shock. Our goal is thus to test whether the post-demand shock demographic composition differs from the status quo. That is, for workers of demographic type $d\in D$, we test whether 
$$ q_{d\cdot}^{\Delta} = \tilde{q}_{d\cdot}$$

We measure $D$ by gender. We consider gender as a key demographic dimension in our context: Men tend to be over-represented among university professors (relative to their share in both the population and the ``pipeline''). In addition, given the demographic composition of habilitants (who are not yet professor), there would be scope for diversification making this a realistic application of the above scenario. We acknowledge though that in general, gender norms were (are) different in East and West Germany. Our research design accounts for this by comparing social sciences and \ac{STEM} in the East net of baseline differences between the subject groups in the West (see discussion in Section~\ref{subsec:academicdesign}). Nonetheless, we show additional results from a second natural experiment to cross-validate conclusions.

\subsection{Results: Changes in the Gender Composition}\label{subsec:genderresults}

Panel~B of Figure~\ref{fig:outc_raw} shows the raw differences in the share of female professors in West Germany (left) and East Germany (right) in 1998. The gray bars indicate the share of female professors in the social sciences, the white bars in \ac{STEM} subjects, respectively. In West Germany, the share of female professors is higher in the social sciences (6.73 percent) than in \ac{STEM} (4.21 percent). In the East, the share of female professors in \ac{STEM} is higher than in the West (6.03 percent), but more comparable in the social sciences (7.34 percent). In other words, the share of female professors in the post-reunification period is considerably higher in \ac{STEM} subjects in the East than in the West, whereas the female share in the social sciences is quite similar in the East compared to the West. This lower (raw) gap in the difference in the share of female professors in the social sciences (affected by the staff replacements) versus \ac{STEM} subjects (unaffected) between East and West Germany suggests a decrease in the share of female professors in the social sciences in the course of the staff replacements. To illustrate this, recall that in Section~\ref{subsec:academicdesign} we derived a benchmark of this raw ``\acl{did}'' of 3.8 percentage points.\footnote{The pre-staff replacements difference in the share of female professors between the social sciences and \ac{STEM} in the East was 9.2 percentage points, in the West 5.4 percentage points.} In 1998, we find a (raw) gap of -1.21 percentage points which is smaller than 3.8 percentage points and even reverses signs.\footnote{The raw gap is the difference in the social sciences versus \ac{STEM} share of female professors in East Germany ($7.34-6.03=1.31$) and the same difference in West Germany ($6.73-4.21=2.52$).} In line with the conclusions above, this suggests that the social sciences in East Germany became more similar to West Germany in terms of the gender composition.

The regression results in column~(4) of Panel~B in Table~\ref{tab:did} confirm the descriptive results. The coefficient is not statistically significant and, if anything, negative. While we cannot reject the hypothesis that the demographic composition was held constant, the results are still suggestive that the demographic composition of the new hires in the social sciences resembled the one of West German professors. An increase in the share of female professors cannot be detected. 

\paragraph{Robustness.} Columns~(5) and~(6) provide the results of robustness tests when adding either cohort fixed effects (birth year) or academic age fixed effects (time since habilitation). These specifications account for the fact that younger cohorts might have a higher share of females and that gender norms could have changed across cohorts. As the replacement hires on average are younger (see Figure~\ref{fig:fs2}), any (potentially) observed changes in the gender composition could be driven by such compositional changes. However, our empirical results hardly change when accounting for such compositional changes. 

\paragraph{Interpretation.} Taken together, our results provide suggestive evidence that open positions during the staff replacements were predominantly filled with men. That is, if anything, the share of the predominant demographic type increases as a result of the labor demand shock. In terms of the framework, this means that while we cannot reject $ q_{d\cdot}^{\Delta} = \tilde{q}_{d\cdot}$ as estimates are imprecise, our point estimates are consistent with $ q_{M\cdot}^{\Delta} \geq \tilde{q}_{M\cdot}$, a (weak) increase in the share of male professors (the predominant demographic type).

\paragraph{Pipeline.} A natural question that follows from this is whether there were, in fact, qualified women to fill the open positions or whether, to the contrary, there was a ``missing pipeline''. Thus, it is difficult to interpret the results above without acknowledging the role of a potentially restricted pipeline of female researchers. To investigate this, we perform several simulations to understand the extent to which a restricted supply of female researchers could have contributed to the above results. 

\begin{figure}[t!]
    \centering
    \includegraphics[width=0.7\linewidth]{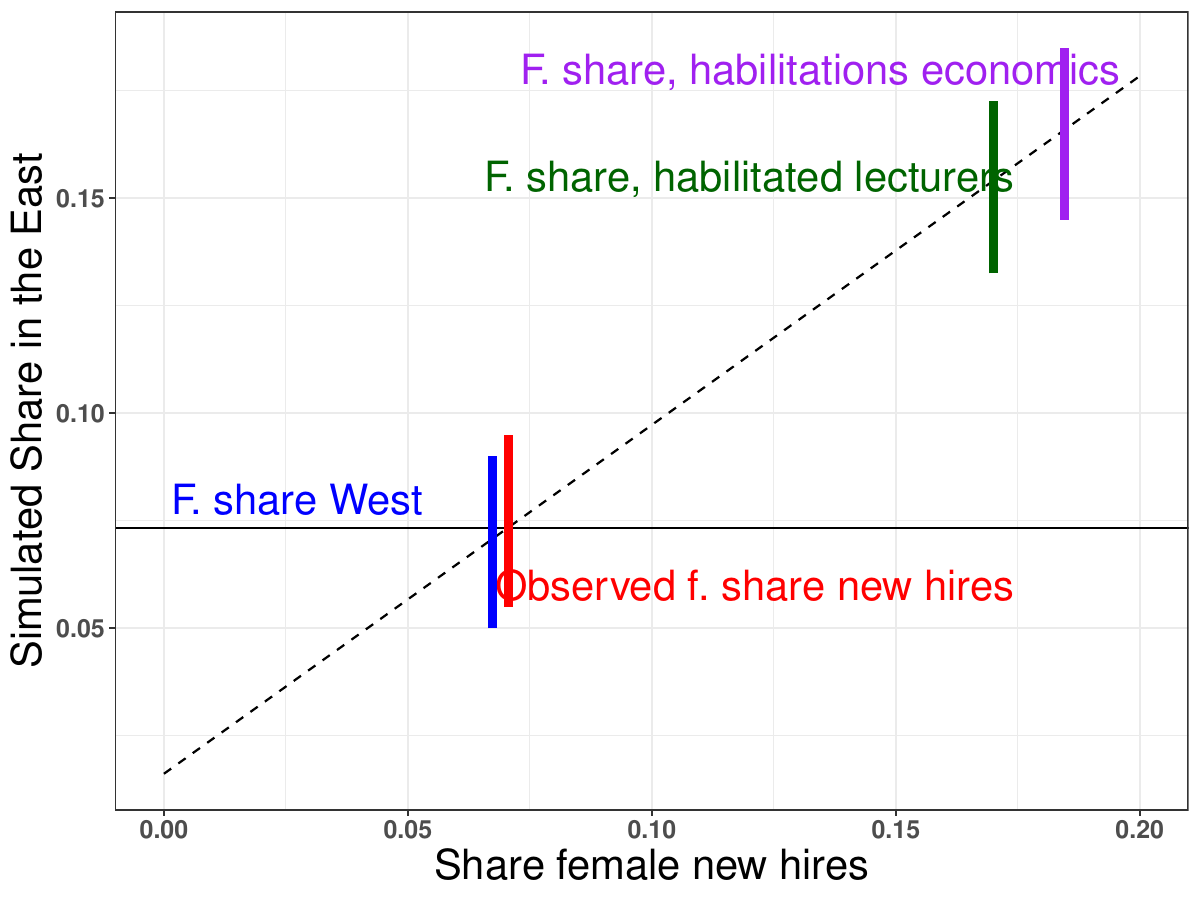}
    \caption{
        \textbf{Simulations of female pipeline.}\\
        \footnotesize{
        \emph{Notes.}- The figure shows the simulated share of female professors in the social sciences in East Germany (in 1998) that would have been observed if the newly appointed professors had been hired according to the proportion shown on the x-axis (dashed diagonal). The horizontal line indicates the observed female share in the East in the Social Sciences. The vertical lines indicate different scenarios on this diagonal. The red vertical line (second from left) indicates the shares observed in the data, the other vertical lines simulated shares. These indicate the simulated share of female professors had replacement occurred according to (i) the observed share among professors in the West in 1998 (blue, first from left), (ii) the share of women among habilitated lecturers in 1998 employed in the Social Sciences in the West (green, third from left), (iii) the share of women among newly habilitated economists 1987-1991 in the West (purple, fourth from left). 
        \emph{Sources.}- Observed shares of professors and habilitated lecturers based on professors and academic staff with habilitation in 1998 in personnel statistics for academic staff \citep{destatis1998personal}, shares among habilitated economists based on \citet{deutschenationalbib}, and based on own calculations.
        }
    }
    \label{fig:simu_gender}
\end{figure}

Figure~\ref{fig:simu_gender} shows what share of female professors would have been observed if hires had been made according to different replacement rates. The red vertical line indicates on the diagonal the observed state (on the x-axis the observed share of females among the new hires, on the y-axis the resulting share of female professors in the social sciences in East Germany). We simulate three benchmarks: First, the blue vertical line highlights the point on the diagonal that represents the share of female professors that would have been observed in the data, had the replacement been performed according to the \emph{contemporaneous} share of female professors in West Germany (not age adjusted). This is almost identical with the actual observed share, even though the recruited professors in East Germany were substantially younger than their West German counterparts (see Figure~\ref{fig:fs2}). Second, the green vertical line highlights the point that represents what one would have observed, had the hiring been done proportionally to the share of habilitated lecturers still working at West German universities. Habilitated lecturers formally qualify for professorships and are thus a relevant approximation to the ``pipeline''. Third, the purple vertical line depicts the point on the diagonal that corresponds to the share that one would have observed had the hiring been proportional to the average female share among newly habilitated economists between 1987-1991. The second and third scenario both would have resulted in a substantially higher share of female professors in the social sciences in East Germany than observed in the data. 

We interpret the above results as suggestive that the low share of female professors among the new hires was not mainly the result of a missing pipeline of (formally) qualified female researchers. Of course, the above analyses cannot account for unobservable differences between formally qualified candidates. Similarly, we cannot observe whether formally qualified female candidates did not want to relocate to an East German university. Again, we stress here that our results do not allow for a conclusion on the barriers on the demand side (labor market discrimination) or on the supply side (mobility restrictions). However, we can conclude that the share of female professors in the social sciences in East Germany after the replacements almost looks like a mirror image of the gender shares at West German universities.

\subsection{Alternative Experiment: University Openings}\label{subsec:uniopenings}

As the German reunification is a unique context, we complement our analysis with a second natural experiment---the university expansion in West Germany during the 1960s and 1970s. Spurred by a scientific rat-race between the global West and the East during the time of the Cold War and during the time period of the \emph{Miracle on the Rhine} (``Wirtschaftswunder''), access to higher education moved into focus of the public debate in West Germany. Increasing demand for higher education, due to generally low education levels, larger birth cohorts and increasing demand for high-skilled work, and investments into high schools added to the pressure on West Germany's higher education system. In 1964, Georg Picht, a prominent German philosopher and educator, coined the term ``educational crisis'' with which he described the lack of higher education opportunities in West Germany at the time \citep{picht1964}. The policy response to the critics and the demands of the time was a reform package modernizing higher education. Importantly, universities were opened where there were none before. Examples include well-known institutions today, such as the universities of Bremen or Bochum. 

The university openings were a large-scale reform that led to a substantial expansion of the university landscape: Compared to the 34 universities in West Germany in 1965, in 1978 there were as many as 63. In other words, the number of universities almost doubled in just over a decade. Figure~\ref{fig:history_uniopen} visualizes the scale of this expansion: Panel~A plots the number of universities in West Germany (black squares) until 1990. There is a fast increase during the second half of the 1960s and the first half of the 1970s, before numbers stabilize during the 1980s. In comparison, the number of universities in East Germany (gray circles) remained almost constant during the same time period. The fast expansion meant that many jobs were created at the new institutions within a relatively short time-frame. Panel~B plots the number of full-time staff at universities in West Germany (gray triangles) next to the dynamics of the university openings (black squares, as in Panel~A). While we do not have data for the 1960s, staff numbers in the 1970s followed the increases in university numbers and rose substantially. The new universities and jobs were not all created in one region. To the contrary, Panel~C illustrates that the new universities were built in regions that had no local access to tertiary education before the reform.\footnote{See \citet{boelmann2023} for further details and for an evaluation of the impact of the university expansion on participation in higher education for more and less mobile population groups.} Given the long qualification period to become professor in Germany compared to the rather short time-frame of the university expansion, the reform plausibly led to a tightening of the academic labor market.

\begin{figure}[t!]
    \centering 
    \includegraphics[width=0.7\textwidth]{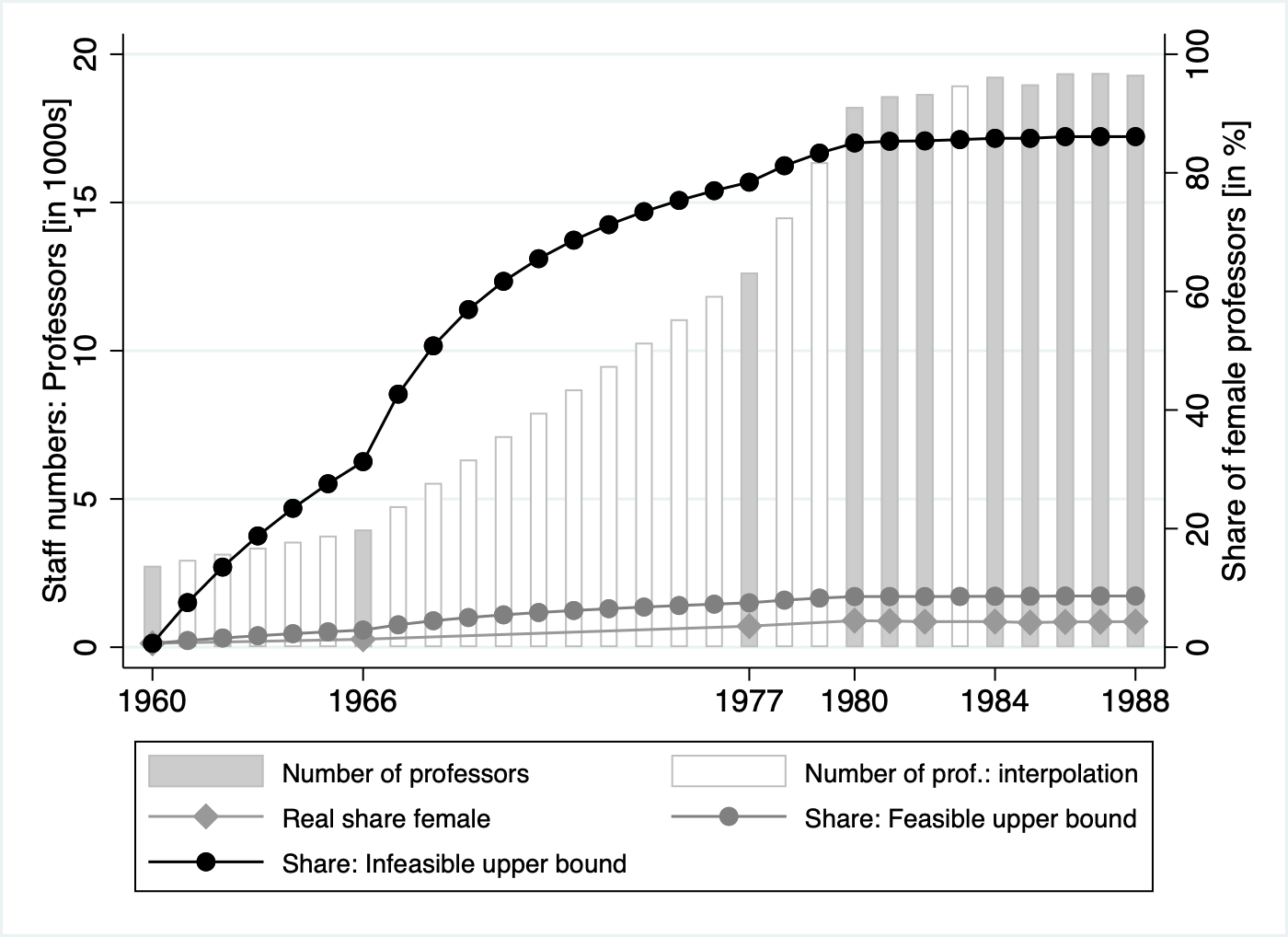}
    \caption{
        \textbf{Female professors in West Germany 1960-1988.} \\
        \footnotesize{
        \emph{Notes.}- The figure shows the observed number (bars) and share (light gray diamonds) of female professors among West German professors from 1960 to 1988. The light gray bars show the actual number of professors employed and the outlines in between show the linear imputation in the number of professors between years. The observed numbers are compared to two (simulated) upper bounds of what would have happened if (i) all openings were filled according to the female share of lecturers in the year before (medium gray circles, feasible upper bound) or if (ii) all openings were filled with only women (dark gray circles, infeasible upper bound). 
        \emph{Sources.}- Based on official statistics by the German Federal Statistical Office---\citet{statistischesbundesamt1960}, Table~1 for 1960; \citet{statistischesbundesamt1966}, Table~A.2 for 1966; \citet{statistischesbundesamt1977}, Table~3.1 for 1977; and \citet{statistischesbundesamt1990}, Table~3.4 for 1980, Table~3.5 for 1981--1988---and own calculations.
    }}
    \label{fig:simshare_uniopen}
\end{figure}

In the absence of individual-level data for this time period, we digitized archival data on university personnel that were compiled and published in statistical reports by the German Federal Statistical Office \citep{statistischesbundesamt1960, statistischesbundesamt1966, statistischesbundesamt1977, statistischesbundesamt1990}. These publications contain aggregate information about university staff at German universities. 
We manually digitized the available years to create a dataset of the share of female university professors in West Germany in 1960, 1966, 1977, 1980-1982 and 1984-1988.\footnote{We transcribed and digitized the data and converted them into csv-files. For the transcription, we used the Software Abbyy Fine Reader that recognizes table structure and reads table contents. We manually cross-checked whether the software transcribed the entries correctly. In some cases, the quality of the historical documents was such that the software could not automatically recognize all rows and columns; in these cases, we made manual adjustments. 
To harmonize the staff categories over the different years, we followed classifications introduced in the official statistics in the 1970s. For the years before 1970 (1960 and 1966), we followed \citet{lundgreen2009} who harmonized these classifications going back until 1953.} 
Figure~\ref{fig:simshare_uniopen} depicts the share of female university professors over this time period. The bars show the observed (gray) and, for the missing years, the interpolated (white) number of female professors. In 1960, there were 18 female professors in total. This number grew to 53 in 1966, 449 in 1977 and 834 by 1988. The gray diamond markers translate these absolute numbers into the share of female among all university professors: In 1960, this share was as low as 0.66 percent, growing to 1.34 percent in 1966, 3.55 percent in 1977 and 4.32 percent in 1988.

As before, an open question is whether one can attribute the low number of female professors to a missing pipeline. This might be particularly relevant for the 1960s and 1970s. To understand the extent to which this is a plausible explanation, we simulated the share of female professors under two scenarios. First, we assume that all new positions were filled with females (black circles in Figure~\ref{fig:simshare_uniopen}). This would have resulted in a share of female professors of over 80 percent. Clearly, this would not have been feasible and constitutes an infeasible upper bound of the share of female professors. Second, we thus assume that the positions were filled according to the female share among lecturers (i.e., formally qualified candidates in the pipeline; gray circles). We call this the feasible upper bound of the female share. In all years, this share exceeds the actually observed share---again suggesting that the pipeline of qualified women would have allowed for a higher share of female professors. Importantly, we again do not make a statement about whether this gap is driven by the supply side (e.g., mobility restrictions for women at the time) or the demand side (e.g., preferences for male candidates).

\section{Trade-Off between Worker Quality and Worker Demographics}\label{sec:tradeoff}

Sections~\ref{sec:workerchars} and~\ref{sec:workerdem} discussed whether the labor demand shock in the aftermath of the reunification induced changes in the workforce composition in terms of (i) (perceived) worker quality and (ii) worker demographics. The results of these analyses indicate that the academic labor market opened up to candidates from lower-ranked universities and that the share of female professors, if anything, weakly decreased. Connecting this with our framework in Section~\ref{sec:model}, we concluded so far that the empirical evidence is consistent with $q_{\cdot H}^{\Delta} < \tilde{q}_{\cdot H}$, a decrease in the share of workers of the (perceived) high-quality type, and $ q_{M\cdot}^{\Delta} \geq \tilde{q}_{M\cdot}$, a (weak) increase in the share of male professors (the predominant demographic type). Combined, these results imply that the data are consistent with what we labeled scenario two earlier: Absent a filled pipeline of high-quality signal workers of predominant demographic (male) type (MH), there is a substitution towards lower-quality signal workers of the same predominant demographic type (ML). In this section, we take this trade-off to the data with the aim to investigate whether indeed $$q_{d^* H}^{\Delta} < \tilde{q}_{d^* H}$$ as implied by the model.

\subsection{Descriptive Evidence}\label{subsec:tradeoff_desc}

We start by looking at whether the relative change in the (perceived) quality dispersion, as discussed in Section~\ref{sec:workerchars}, is the same for males and females. To do so, we compute the dispersion measure described above separately by gender.\footnote{Since women faced higher barriers in academia, they are likely positively selected in terms of inherent quality. If hiring committees apply stricter standards to women, departmental prestige accurately measures women's quality relative to other \emph{women} but may understate it relative to \emph{men}. Additionally, a glass ceiling that concentrates women at lower-ranked departments could mean that the smaller widening in female dispersion reflects pre-existing bunching rather than a genuine quality differential, in which case we would underestimate the true quality downgrade associated with hiring more women from the same departments. Neither concern affects our main results, however, as the increase in dispersion is driven entirely by the hiring of additional male professors: we find no effect of the labor demand shock on the hiring of female professors.} Using this gender-specific dispersion index, we augment our baseline specification in equation~\eqref{eq:exp2_did1} to allow the effect to vary by gender. The results are shown in column~(3) of Panel~A in Table~\ref{tab:did}. Conditional on everything else, we find that the increase in dispersion in academic backgrounds is by about~27\% lower for female than for male professors. However, there is a lack of precision and the estimated coefficient for the triple interaction term is not statistically significant.

\begin{figure}[p]
    \centering
    \includegraphics[width=\textwidth, height=0.8\textheight, 
                     keepaspectratio, trim=0 0 0 0, clip]{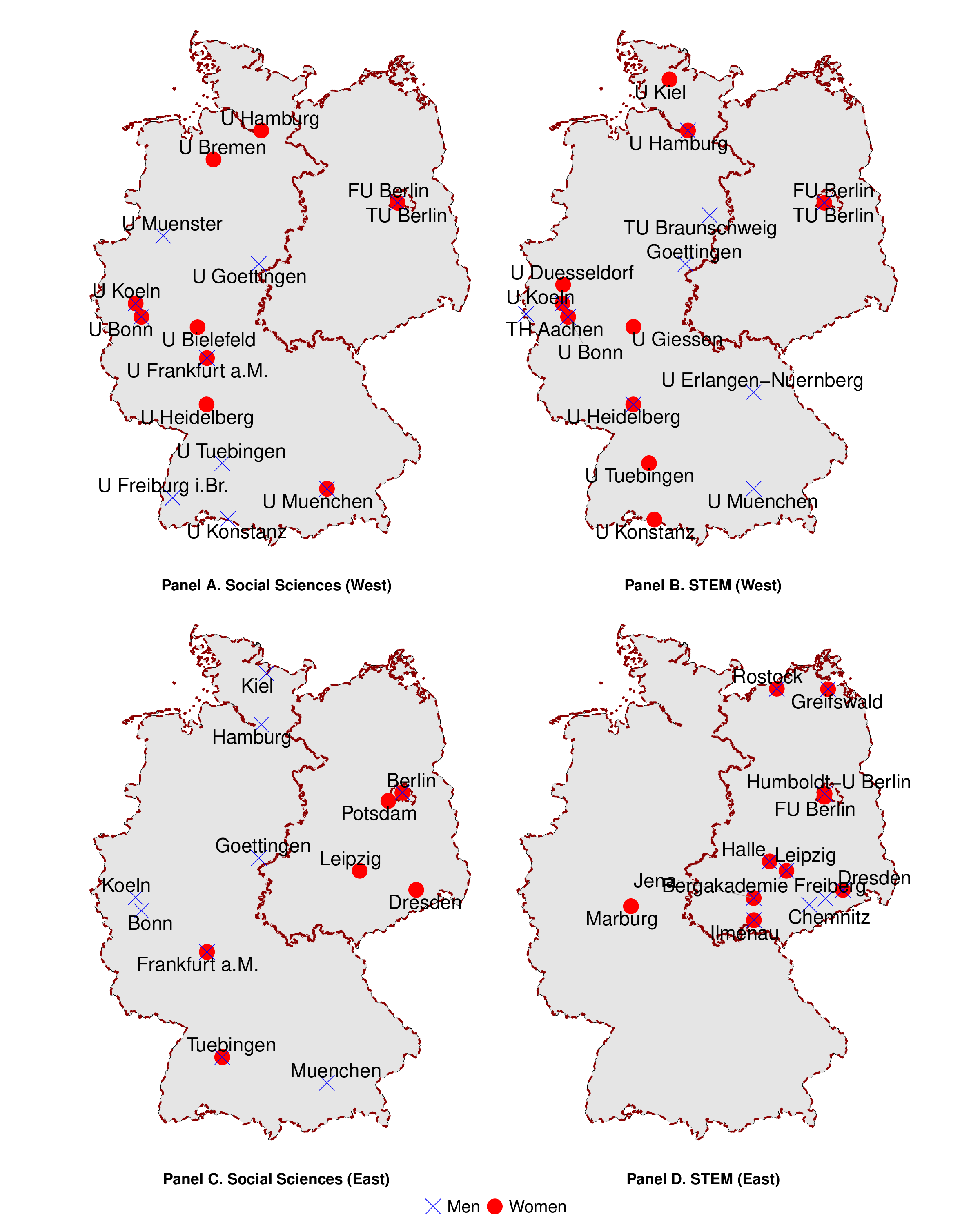}
    \caption{
        \textbf{Ten most common habilitation departments of professors in 1998.} \\
        \footnotesize{
        \emph{Notes.}~The figure shows the locations of the ten most common habilitation 
        departments of professors employed in 1998 in West Germany (top, Panels~A and~B) 
        and in East Germany (bottom, Panels~C and~D), for the social sciences (left, 
        Panels~A and~C) and \ac{STEM} (right, Panels~B and~D). The most common habilitation 
        departments are shown separately by gender: Blue crosses represent the habilitation 
        departments of male professors, red dots those of female professors. The dashed line 
        within Germany represents the former inner-German border.
        \emph{Sources.}~Personnel statistics for academic staff 
        \citep{destatis1998personal}, professors in 1998. German boundaries taken from \citet{gmu_germany_coldwar_boundary}. 
        }
    }
    \label{fig:top10_maps}
\end{figure}

To thus look at this question more descriptively, we study the ten most common universities from which male and female professors in 1998 had obtained their habilitation---both in East and West Germany and separately for social sciences and \ac{STEM}. Figure~\ref{fig:top10_maps} illustrates the patterns. Panels~A and~B show the ten most common habilitation universities for West Germany (red circles for female professors, blue crosses for male professors), Panels~C and~D for East Germany. Professors in ~\ac{STEM} subjects in West Germany in 1998 largely obtained their habilitation from West German universities and vice versa for those in the East. This pattern is comparable for male and female professors. For social sciences, the picture looks different: Professors in West Germany in 1998 obtained their habilitation predominantly from West German universities. The same holds for male professors in East Germany in 1998---but not for females. Together with the regression results above, this strongly suggests that the newly opened positions were filled with male candidates from West Germany who were hired from (also) lower-ranking departments. 

\subsection{Evidence from Simulations} 

To go beyond these descriptives and to make an attempt at translating our results into quality losses, we turn to simulation analyses. Our simulation exercise aims at providing bounds for the quality losses incurred by keeping the predominant demographic type (here: males) at least constant. We do so under varying assumptions regarding the potential applicant pool to explore counterfactual ``what if'' scenarios with respect to the average quality of newly hired professors. The advantage of these simulations is that they allow us to model the otherwise unobserved (latent) quality of the workers. We proceed in three steps.

In the first step, we determine the size of the potential pool of male and female applicants, respectively, by using the total number of habilitation graduates from West German universities in the social sciences in the years 1987-1991 \citep{deutschenationalbib}.\footnote{Given that newly hired professors mainly received their habilitation in West Germany (compare Figure~\ref{fig:fs1}), we use the graduates from only West German universities as potential pool of applicants for our simulation exercise.} In total, there are 990~female and  4377~male graduates who received their habilitation directly prior to reunification and can thus reasonably be expected to  be available on the academic labor market at the time of the staff replacements. 

Step two assigns to each male and female in the applicant pool a simulated quality. This can be thought of as a latent measure of quality that encompasses various aspects of academic jobs, including research and teaching. Thus, we assume that quality can be summarized in a single index, which we draw from a standard normal distribution. We simulate two different scenarios in which we draw the quality for both male and female candidates using random standard normal distributions under different assumptions. Specifically, we simulate:
\begin{itemize}
    \item[1.] Conservative scenario: The quality distribution of the potential pool of applicants is identical for both male and female habilitation graduates, i.e., $Q_M\sim\mathcal{N}(0,1)=Q_F\sim\mathcal{N}(0,1)$. We call this scenario conservative, because even though the pool of potential female applicants is much smaller, there is no positive selection into the applicant pool, i.e., women do not positively select into a habilitation. Note that this contradicts previous findings on women in academia in the 20th century \citep{iaria2024}.
    \item[2.] Non-conservative scenario: Women in the pool are positively selected for quality. We draw the male applicants' pool quality as before, i.e., $Q_M\sim\mathcal{N}(0,1)$, and then assign the best female applicant the same quality as the best male applicant, the second-best female applicant the same quality as the second-best male applicant, etc. 
\end{itemize}

In the third step, we then determine the observed gender-specific quality hiring cut-off, based on the observed number of male and female professors in East Germany post reunification. This means that we take the observed number of male and female professors in the social sciences in the East as given, and calculate quality thresholds for the marginal hire. For this, we assume merit-based hiring within gender, i.e., the best male (and female, respectively) applicant is hired first, then the second-best, etc.. Thus, the quality threshold can be interpreted as the ``quality'' level of the last hire for both men and women. 

\begin{figure}[h!]
    \centering
    \makebox[0.5\textwidth][l]{\footnotesize{\textbf{Panel A.} Conservative scenario -- women}}%
    \makebox[0.5\textwidth][l]{\footnotesize{\textbf{Panel B.} Conservative scenario -- men}}\\[2pt]
    \includegraphics[width=0.5\textwidth]{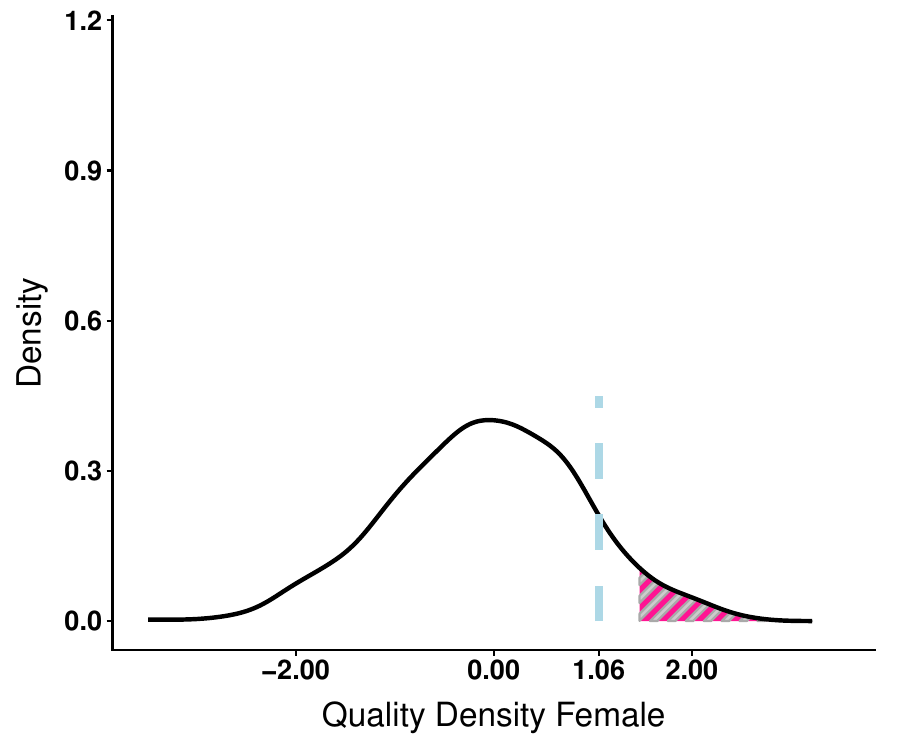}%
    \includegraphics[width=0.5\textwidth]{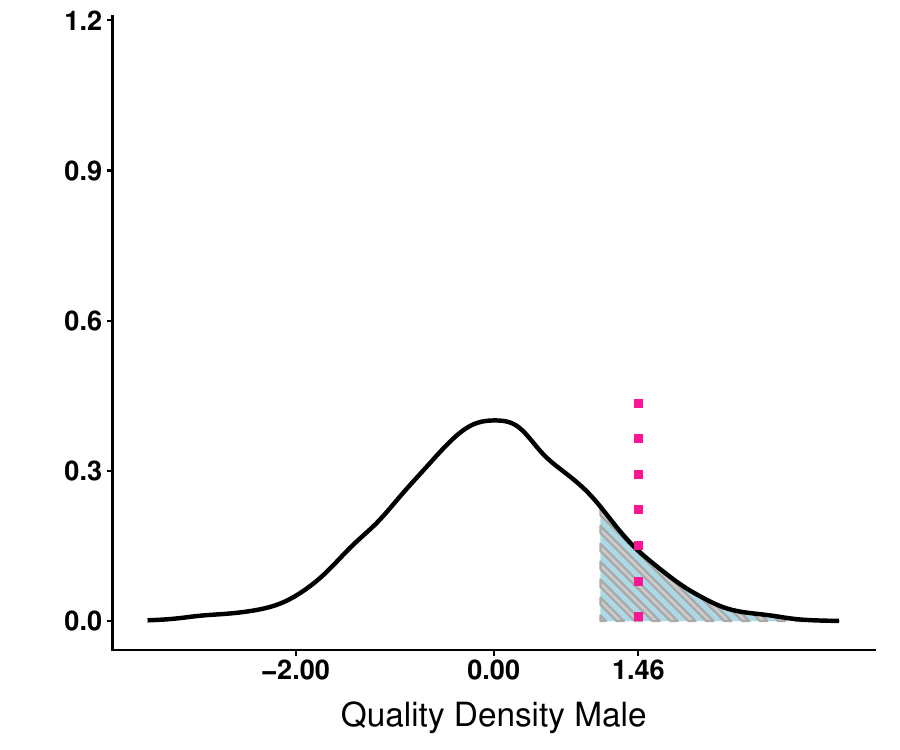}\\[6pt]
    \makebox[0.5\textwidth][l]{\footnotesize{\textbf{Panel C.} Non-conservative scenario -- women}}%
    \makebox[0.5\textwidth][l]{\footnotesize{\textbf{Panel D.} Non-conservative scenario -- men}}\\[2pt]
    \includegraphics[width=0.5\textwidth]{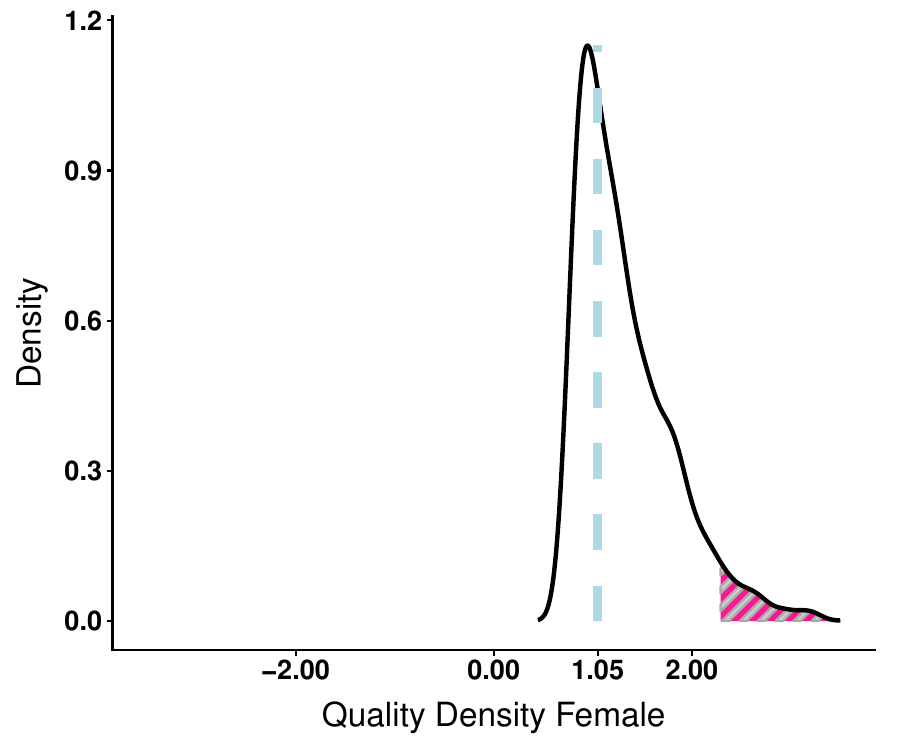}%
    \includegraphics[width=0.5\textwidth]{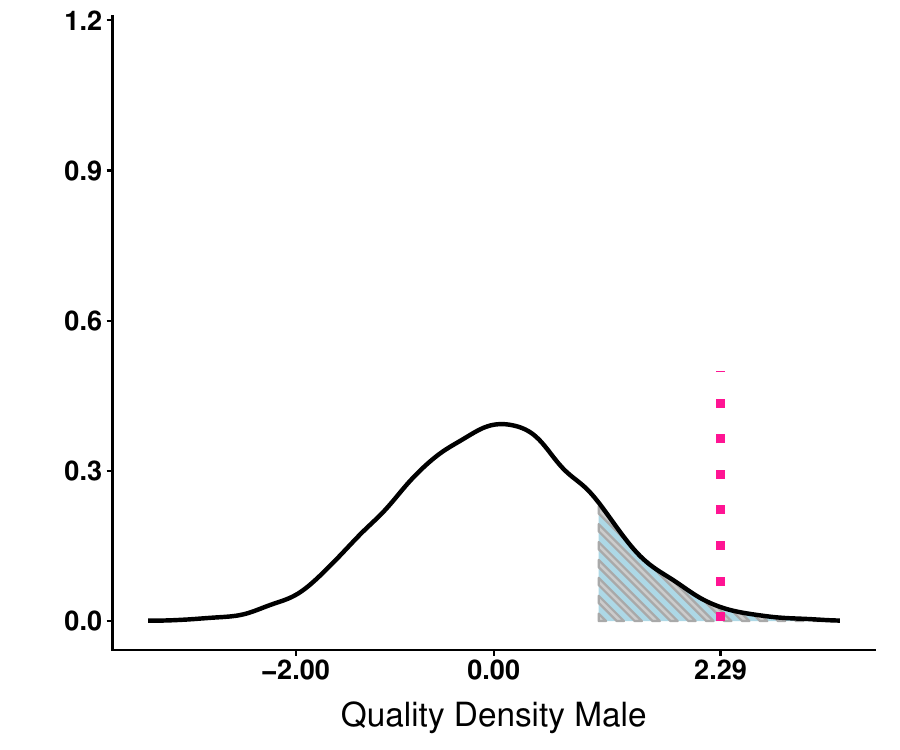}
    \caption{
        \textbf{Simulated quality thresholds for male and female hires.} \\
        \footnotesize{
        \emph{Notes.}~The figure shows the quality distribution in the female applicant pool 
        (left, Panels~A and~C) and in the male applicant pool (right, Panels~B and~D). The 
        shaded area represents the quality until the last hired applicant from the same gender, 
        the vertical line the quality cut-off for the other gender. The gap between the shaded 
        area and the vertical line is the implied quality difference between the last female 
        applicant hired and the last male applicant hired. Panels~A and~B represent the 
        conservative scenario, where quality distributions for male and female applicants are 
        identical. In Panels~C and~D, we assume that the female pool is positively selected 
        for quality.
        \emph{Sources.}~Numbers of male and female hires based on personnel statistics for 
        academic staff \citep{destatis1998personal}, professors social sciences in 1998 in 
        East Germany, numbers of the male and female pool size based on \citep{destatis1998personal}, and own 
        calculations.
        }
    }
    \label{fig:sim_quality}
\end{figure}

The results of these simulations are shown in Figure~\ref{fig:sim_quality}. In the conservative scenario (Panels~A and~B), the quality cutoff for male applicants (as indicated by the vertical blue line in Panel~A) is 1.06, for female applicants 1.45 (vertical pink line in Panel~B). This means that the last male candidate hired is positioned close to 0.5~of a standard deviation lower in the quality distribution than the last female candidate hired. In the non-conservative scenario (Panels~C and~D), where we allow for a positive selection of women into the applicant pool, the gap in the simulated quality thresholds becomes even more pronounced: In this scenario, the threshold for male candidates is 1.05 but 2.29 for female candidates. This results in the last male hire being 1.2~standard deviations lower in the quality distribution than the last female hire.

Finally, we examine how negatively selected the pool of potential female professors would have had to be in order to reach quality parity of the last male and female hire in a gender-blind, merit-based hiring regime. To do this, we keep the (simulated) quality of men in the applicant pool fixed while varying the quality of women until we reach the same  threshold for both men and women (recall that above, the threshold was higher for females). First, we simulate the gender-specific quality distributions in the population, from which we assume the applicant pools are drawn. In general, we assume that both male and female candidates for professorships are positively selected from the population in terms of our quality measure, i.e., the applicant pools are drawn from the right-hand tail of this distribution (for instance, from the 95th percentile or top five percent). Second, we start with the conservative scenario from above as the benchmark for the following simulations. Recall that, in the conservative scenario, we assume the pool of both male and female candidates to have the same \emph{average quality}\footnote{We draw applicants from the 95th percentile of the population quality distribution. This threshold is arbitrarily (conservatively) chosen and reflects the assumption that in general, academics are positively selected from the population. Varying over the benchmark percentile does not affect the qualitative results, but it matters for the quantitative result, as we discuss below.}, resulting in different quality thresholds for men and women.

 
Third, to determine the point where negative selection of female candidates rationalizes the observed gender-disparate hiring in response to the labor demand shock, we draw male candidates from the same fixed high-quality pool (above the 95th percentile of a standard normal distribution) but iterate the female candidate pool quality over 30 different scenarios ranging from being above the 95th to above the 80th percentile. For each of these scenarios, we draw samples of size $n_1=4377$ (male pool) and $n_2=990$ (female pool) from truncated normal distributions and simulate a hiring process where male and female candidates are hired according to the observed numbers in the social sciences in East Germany after the staff replacements. As before, we assume that hiring is merit-based and positions are filled starting at the top of the quality distribution. We are most interested in the hiring threshold, i.e., the quality level of the marginal (last) hire. We calculate this threshold separately by gender for each scenario. By iteratively decreasing the average quality of the pool of female candidates while holding the pool of male candidates constant, we can trace out how negatively selected the female pool would have had to be to result in equal quality of the last male versus female hire. 

\begin{figure}[t!]
    \centering 
    \begin{tabular}{cc}
        \multicolumn{1}{l}{\footnotesize{\textbf{Panel A.} Distribution of simulated quality}} 
        & \multicolumn{1}{l}{\footnotesize{\textbf{Panel B.} Quality cut-offs}}\\
        \includegraphics[width=0.5\textwidth, trim={0 0 0 1.5cm},clip]{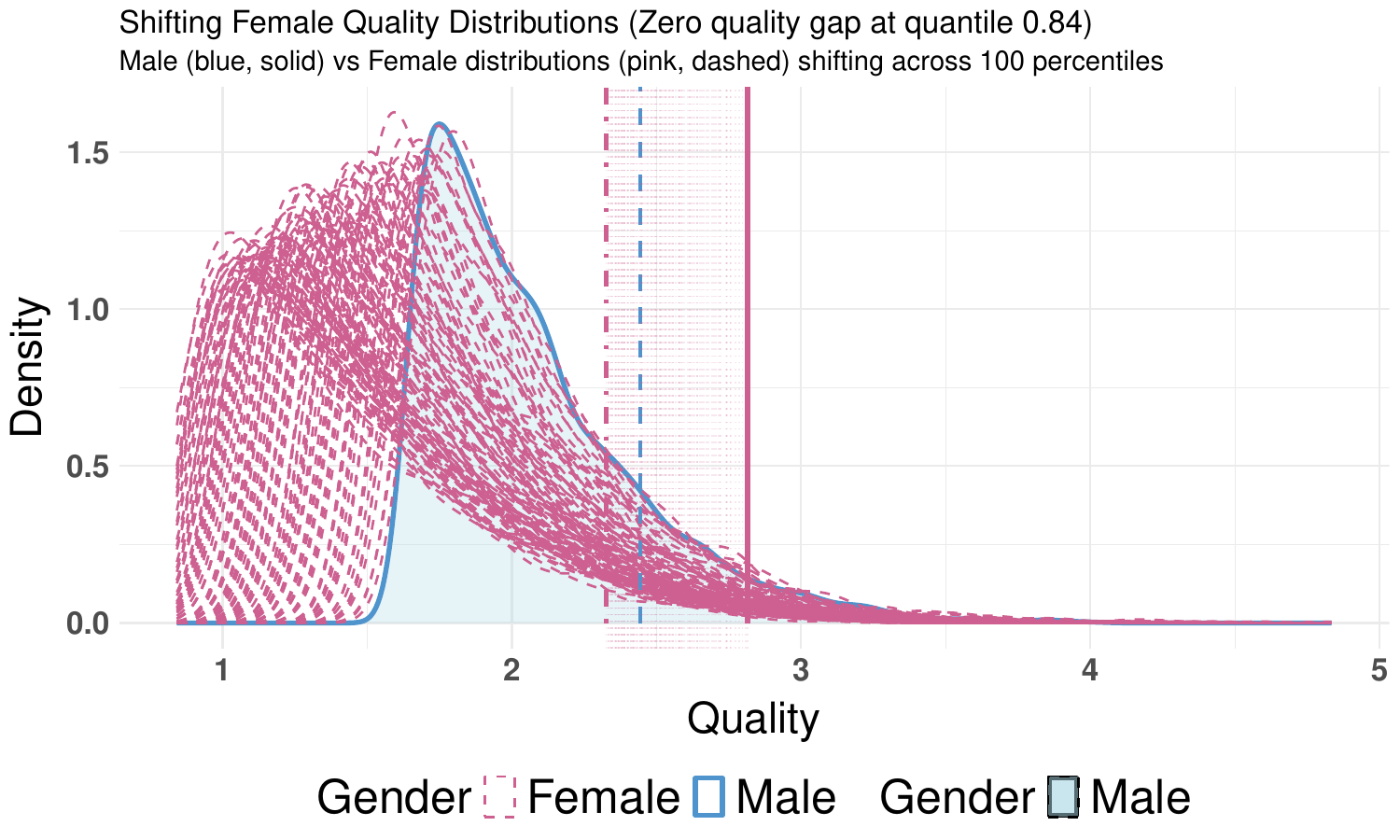  } 
        & \includegraphics[width=0.5\textwidth, trim={0 0 0 1.5cm},clip]{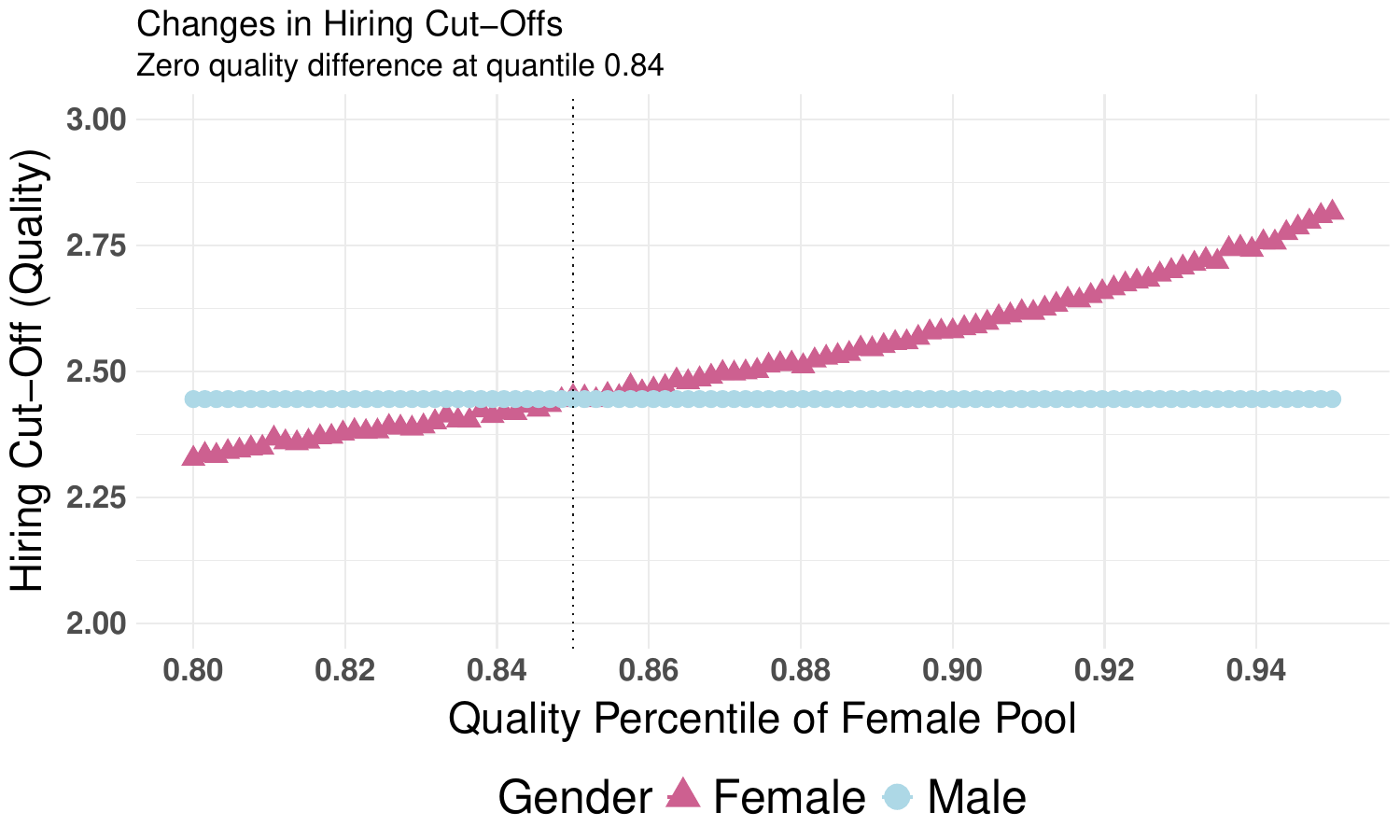} 
    \end{tabular}    
    \caption{
        \textbf{Simulated quality differences with negative selection.} \\
        \footnotesize{
        \emph{Notes.}- The figure illustrates how progressive negative selection into the female applicant pool affects the quality of the last hired woman relative to the last hired man. The male hiring pool is kept fixed (drawn from the 95th percentile of the distribution). Panel~A shows the quality density of the shifting female applicant pool in pink and the fixed male quality density in blue. The dashed blue line marks the quality of the last hired man. The solid pink line marks the quality of the last hired woman in the conservative scenario, where the male and female pools have equal average quality. The dotted vertical lines trace how the quality of the last hired woman shifts as the female applicant pool becomes more negatively selected, i.e., the women in the applicant pool are drawn from increasingly lower percentiles. Panel~B plots this last-hire quality for women (pink line) against the percentile from which the female applicant pool was drawn, holding the male pool quality fixed (blue line). The two coincide at the 85th percentile, where the last hired woman matches the last hired man in quality.
        \emph{Sources.}- See Figure~\ref{fig:sim_quality} and own calculations.
    }}
    \label{fig:sim_quality_II}
\end{figure}

Figure~\ref{fig:sim_quality_II} presents the results. Panel~A visualizes the simulated quality distribution for men (blue, shaded area) and women (pink, dashed lines). Vertical bars show the respective quality cut-offs for filling the positions for men (blue, dashed vertical line) and women (pink, dot-dash and solid vertical lines). The solid pink vertical line depicts the conservative scenario: here, male and female candidates are both drawn from the top five percent of the population distribution. This yields a 0.5~standard deviations lower quality cut-off for male compared to female candidates, as discussed above (see Figure~\ref{fig:sim_quality}).\footnote{Note that this holds regardless of which percentile is chosen as the benchmark, i.e., whether we use the 95th percentile/top five percent or vary this cutoff.} The dot-dash pink vertical line illustrates the quality threshold for females when their pool of candidates is negatively selected compared to males at the lowest percentile considered, with the faint dotted pink lines tracing the thresholds across all intermediate percentiles.

To more easily determine the point at which the negative selection implies the same quality cutoffs for the hired candidates, Panel~B shows the resulting quality thresholds for each of the simulated pools of female applicants (pink triangles, dashed line) and the fixed pool of male applicants (blue circles, solid line). The simulations iterate through the percentiles of the quality distribution that demark the selection of the female applicant pool (x-axis). For instance, a value of 0.8 means that the female applicants are drawn from the top 20 percent of the population distribution. The quality thresholds for females and males cross at the 85th percentile, i.e., when the female pool of applicants is drawn from the 85th percentile and above. Put differently, for the last male and female hire to have equal quality, the pool of women (with a habilitation) would have to be selected from the top 15 percent of the population quality distribution, if the pool of men (with a habilitation) were selected from the top five percent.\footnote{Again we highlight that choosing a different benchmark percentile does not qualitatively change the results. However, less select male applicants pools require increasingly more negatively selected female applicant pools to achieve quality parity for the marginal (last) hire. If we assumed, for instance, that the male candidate pool were selected from the top 20 percent instead of the top 5 percent, we would need to assume that women were selected from the top 62 percent to reach quality parity for the last hire.} 

\subsection{Performance Implications}\label{subsec:implications}

The empirical analysis in Sections~\ref{sec:workerchars} and~\ref{sec:workerdem} has demonstrated that the staff replacements in the social sciences in East Germany, a large labor demand shock, led to the hiring of predominantly younger men from a wider, less positively selected range of West German universities. The simulation exercise has illustrated the sizable implied quality differences between the marginal male and female candidate hired. The results are consistent with the stylized framework and indicate a scenario in which large labor demand increases result in lower quality of the workforce when the applicant pool is fixed and the demographic composition of the workforce is (at least) held constant.

Taking our results at face value, a final question is whether patterns above regarding the trade-off between worker quality and demographics indeed resulted in observable performance reductions in the social sciences in East Germany compared to West Germany. Unfortunately, we cannot answer this question empirically in a satisfactory way. This is for two reasons. First, the job tasks and professional environments of the new professors in East Germany were very different than those in West Germany. Whole departments were restructured and rebuilt, meaning that a lot of time and resources needed to be spent on this, whereas West German professors could build on existing and well-established structures. These differences make it hard to find a comparable performance measure. Second, academic performance is inherently difficult to measure even if job conditions were comparable. One reason for this is that academic performance measures partially depend on networks and reputation. For example, empirical evidence has revealed a large home bias in the publication process \citep{bethmann2023}. Professors in East German social sciences departments might have been at a disadvantage in this respect. Overall, the empirical study of performance consequences is thus extremely difficult if not impossible in our context.  

We thus restrict ourselves to some descriptive evidence for performance implications. Specifically, we consider how departments perform in the university ranking provided by the \acf{CHE}. The ranking puts a special emphasis on various dimensions of academic performance, including research and teaching. For that reason, it is based on observed data, expert evaluations and student surveys. We use the earliest available wave of data, 2005 and 2007, for which we can observe several subjects in both \ac{STEM} and social sciences. We select a range of evaluation criteria to compare averages across fields and across East and West Germany.\footnote{Because of the low number of observations, we refrain from more sophisticated statistical analyses.} Appendix Table~\ref{tab:che_means} shows the differences in differences in raw means. For research (Panel~A), external funds per researcher stand out as the only positively signed difference: The gap between Social Sciences and \ac{STEM} is considerably smaller in the East than in the West. This could be for a variety of reasons, including targeted funding for East Germany. For publications and completed PhD theses, there is if anything a negative difference-in-difference. Student evaluations (Panel~B) are very similar between both social sciences and \ac{STEM} and the East and the West, likely due to the fact that almost all universities in Germany are public and hence subject to roughly equivalent teaching requirements and funding criteria.


\section{Conclusion}

This paper examines how labor demand shocks affect workforce diversity in the absence of targeted diversity policies. We exploit a unique natural experiment arising from German reunification, which created a substantial labor demand shock in East German academia when staff deemed unfit for Western science were replaced. Importantly, these shocks varied by field: while \ac{STEM} disciplines retained most staff, nearly all social sciences positions were vacated. Using a regional \acf{did} approach comparing social sciences (treatment) to \ac{STEM} (control) across East and West Germany, we isolate the causal effect of this demand shock on the workforce composition of professors.

Our analysis of administrative data covering all German professors in 1998 reveals that quality, as proxied by university of habilitation, declined with the demand shock, with male hires in East German social sciences coming from less select departments. In addition, we find that the large labor demand shock did not increase workforce diversity in terms of gender. Despite a pipeline where women comprised 13-20\% of qualified candidates, the replacement process resulted in female representation in East German social sciences converging to the lower West German levels rather than improving. We show that low female shares cannot be attributed to a shortage in ``the pipeline''---the pool would have supported at least twice the observed female representation. We find similar patterns during West Germany's university expansion of the 1960s-1970s, suggesting these dynamics are not unique to the reunification context.

We interpret these findings through a framework where hiring decisions involve trade-offs between demographic characteristics and quality signals, and our evidence is in line with a scenario where the demographic composition is kept constant at the expense of worker quality. Through simulation analysis exploiting pipeline data, we show that hiring quality thresholds were substantially higher for women than for men. Under plausible assumptions about positive selection of women into the academic pipeline, we find that the lowest-quality male hire has a 1.2~standard deviation lower quality level than the lowest-quality female hire. In contrast, when allowing for negative selection of women into the pipeline, we find that rationalizing the observed gender-disparate hiring outcomes in a purely merit-based system would require female candidates to be drawn from the top 15\% of the population and male candidates from only the top 5\%.

We conclude that market forces alone may not be enough to eliminate non-diverse hiring outcomes despite substantial quality losses. While we cannot distinguish whether our results are driven by the demand side (e.g., discrimination), the supply side (e.g., West German women's unwillingness to move) or both, they nevertheless clearly highlight a structural issue with sizable costs to (academic) production. As such, our results contribute a new perspective to the current policy debate around the elimination of \ac{DEI} programs and suggest that targeted policy interventions are likely needed if workforce diversity is an explicit goal. And even if it is not, they might still be beneficial in terms of productivity.



\singlespacing


\printbibliography


\newpage
\singlespacing

\section*{A. Appendix: Additional Figures and Tables}

\renewcommand{\thefigure}{A.\arabic{figure}}
\renewcommand{\thetable}{A.\arabic{table}}
\setcounter{figure}{0}
\setcounter{table}{0}



\begin{figure}[h!]
    \centering 
    \includegraphics[width=1.02\textwidth]{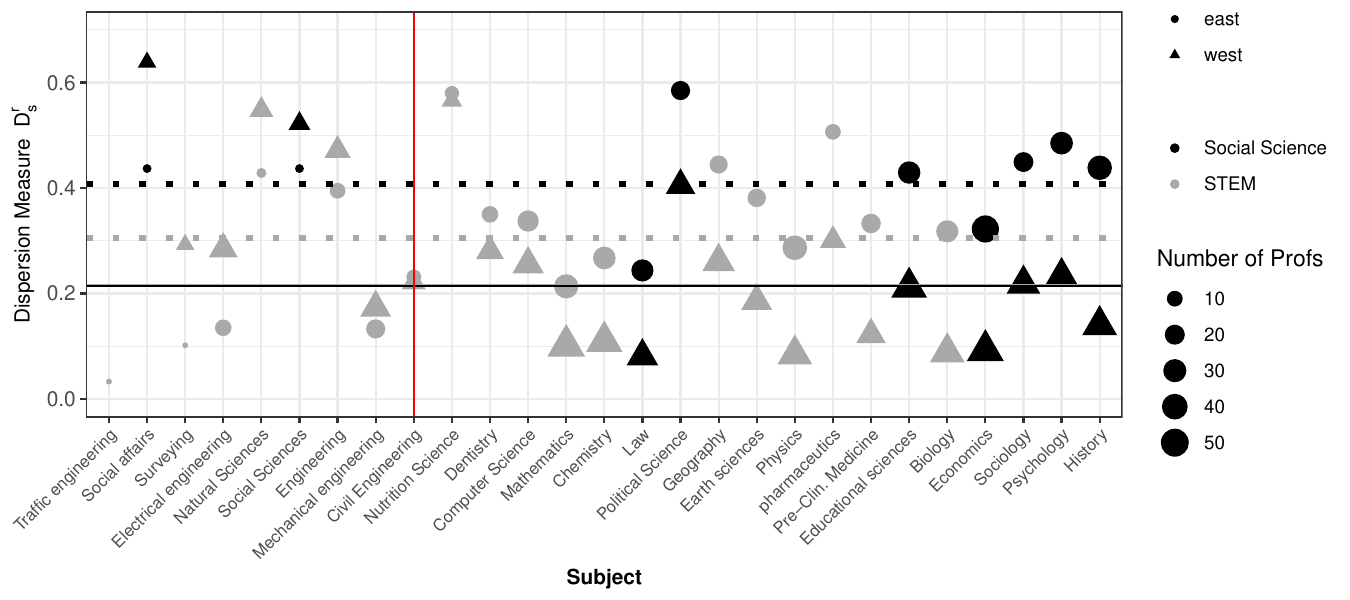}
    \caption{
        \textbf{Dispersion index by subject.}\\ 
        \footnotesize{
        \emph{Note.}- This figure plots the values of the dispersion index from equation~(\ref{eq:index}) by subject, separately for professors in East Germany (circles) and West Germany (triangles) in 1998. The size of the markers corresponds to the number of observations. Social Science subjects are shown in black, \ac{STEM} subjects in gray. The dotted (solid) lines show the weighted means of the dispersion measure in the East (West), for \ac{STEM} (gray) and Social Sciences (black). The subjects are ordered according to the difference between the dispersion measure within the subject between East and West, $\Delta_E=D_s^{E}-D_s^{W}$. If this value is negative, it indicates that the dispersion measure is larger in the West than in the East. Every subject to the right of the red vertical line has a larger dispersion in the East than in the West. This is the case for the majority of subjects in the social sciences. 
        \emph{Sources.}- Personnel statistics for academic staff \citep{destatis1998personal}, professors in 1998, and own calculations.
    }}
    \label{fig:disp_subj}
\end{figure}
\clearpage

\begin{figure}[t!]
    \centering 
    \begin{tabular}{cc}
        \shortstack[l]{
            \footnotesize{\textbf{Panel A.} Universities}\\
            \includegraphics[width=0.45\textwidth]{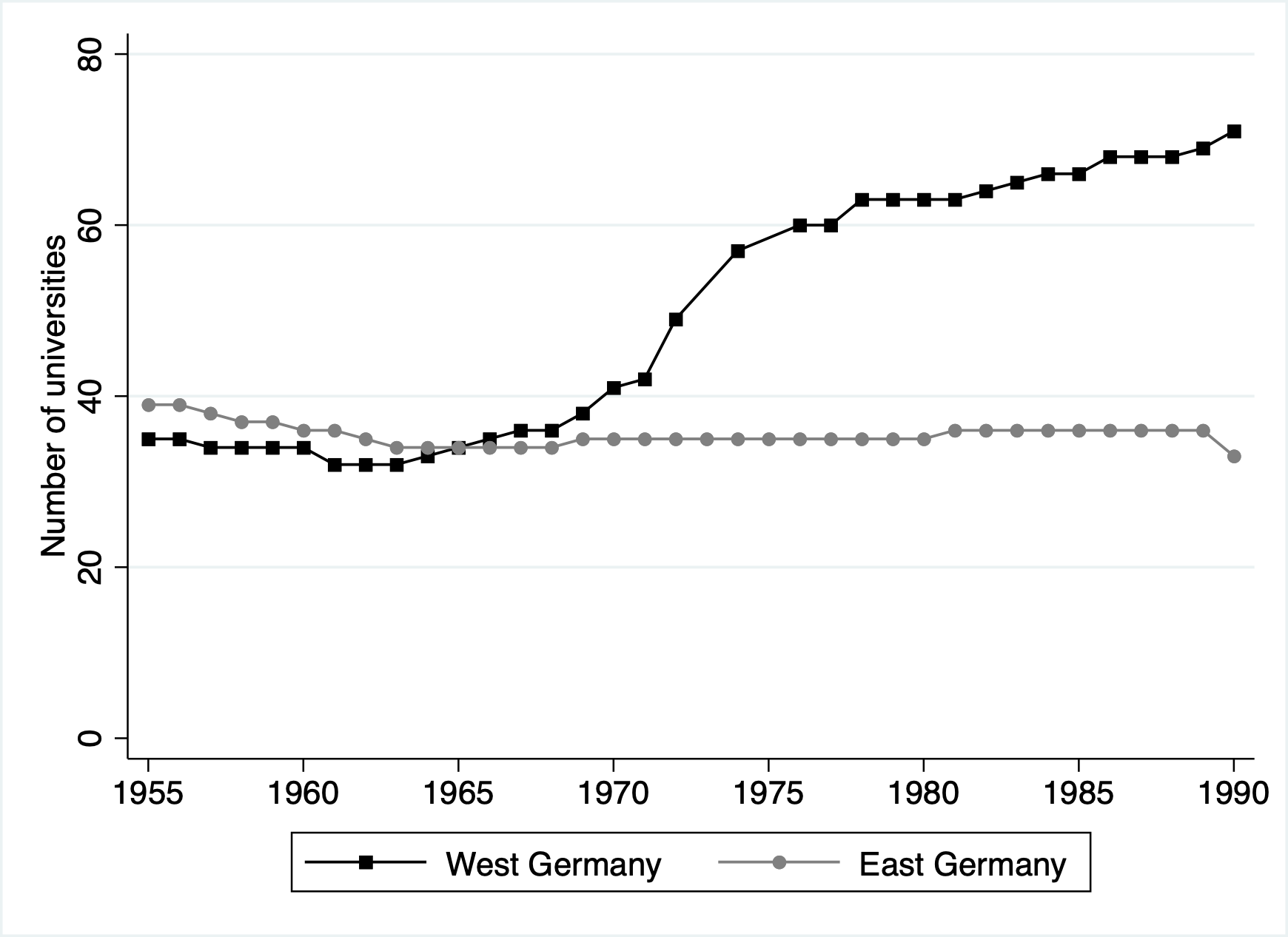}\\
            \footnotesize{\textbf{Panel B.} Full-time university staff}\\
            \includegraphics[width=0.45\textwidth]{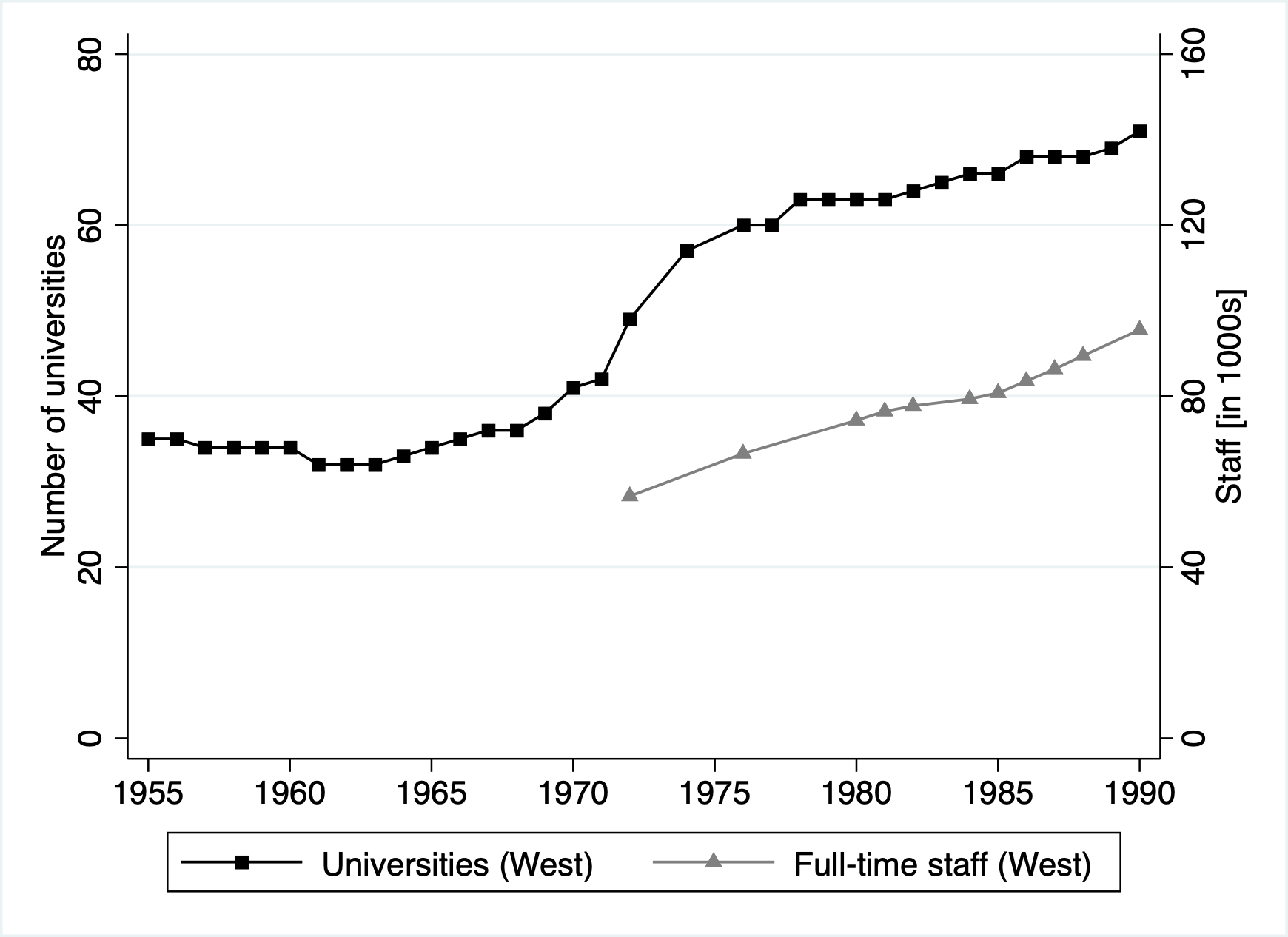}
        }
        &
        \shortstack[l]{
            \footnotesize{\textbf{Panel C.} Geography of university openings} \\
            \includegraphics[width=0.455\textwidth]{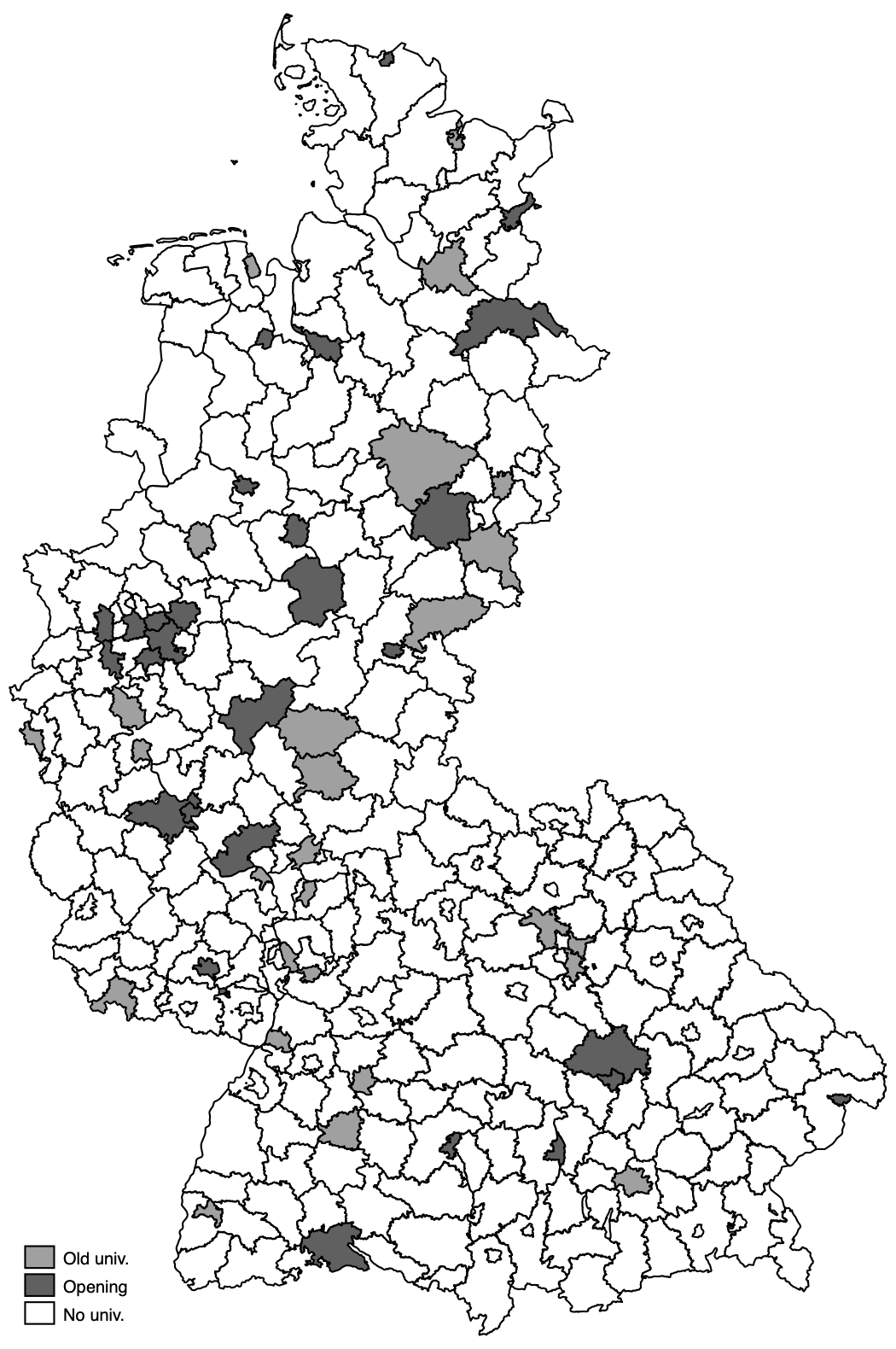} 
        }    
    \end{tabular}    
    \caption{
        \textbf{University expansion in West Germany.} \\
        \footnotesize{
        \emph{Notes.}- Panel~A shows the number of universities according to administrative records in the former West/\ac{FRG} (black squares) and East/\ac{GDR} (gray circles), from 1955 to 1990. Panel~B plots the number of universities in the \ac{FRG} as in Panel~A against the number of full-time staff at universities in the \ac{FRG}, scaled in thousands (gray triangles, selected years only). Panel~C visualizes the geography of the university openings by district: districts with no university (white), with an old university (light gray) and with a university opening between 1945 and 1990 (dark gray).
        \emph{ Sources.}- Panel~A is based on \cite{koehler2014} for the East and \citet{bildungzahlenspiegel, statistischesamtdesaarlandes1952, statistischesbundesamt1953} for the West, and own calculations. Panel~B additionally uses data from \citet{franzmann2006,statistischesbundesamt1990} for staff numbers. Panel~C taken from \citet{boelmann2023}.
    }}
    \label{fig:history_uniopen}
\end{figure}
\clearpage

\begin{table}[t!] 
     \centering
     
\begin{tabular}{lrrrr}
\toprule
&\multicolumn{1}{c}{(1)}&\multicolumn{1}{c}{(2)}&\multicolumn{1}{c}{(3)}&\multicolumn{1}{c}{(4)} \\[1.5ex]
&\multicolumn{2}{c}{\textbf{Ph.D.}} & \multicolumn{2}{c}{\textbf{Habilitations}} \\
\textbf{Year}& \multicolumn{1}{c}{\textbf{N}} & \multicolumn{1}{c}{\textbf{\% women}} & \multicolumn{1}{c}{\textbf{N}} & \multicolumn{1}{c}{\textbf{\% women}} \\ 
\midrule
 1976  &  701 &21.4 & 38&8.2\\
 1977 &645 & 22.5 & 49& 10.8\\
 1978 & 795 & 25.4 & 55 &10.0 \\
 1979 &830 & 27.1 & 61  & 9.3\\
 1980 & 1.007 & 30.1 & 97 & 13.7\\
 1981 & 1.044 & 31.6 & 86& 13.3\\
 1982 & 1.182 & 33.5 & 76 & 11.5\\
1983 & 1.361 & 33.4 & 76 & 10.3\\
 1984 & 1.571 & 34.7 & 98 & 11.8\\
 1985 & 1.851 & 32.8 & 112 &  13.2\\
 1986 & 1.805 & 31.8 & 138 &  14.9\\
\bottomrule
\end{tabular}
     \caption{
        \textbf{Junior researchers in East Germany 1976-1986.}\\
        \footnotesize{
        \emph{Notes.}- This table presents the number and share of female students who received a PhD (columns~1 and 2) and a habilitation (columns~3 and 4) in the \acf{GDR} between 1976 and 1986. 
        \emph{Sources.}- \citet{budde2003frauen}. 
        }
    }
    \label{tab:diss_women_gdr}
\end{table}


\begin{table}[b!] 
     \centering
     {\normalsize
\begin{tabular}{llrr}
\toprule 
\addlinespace[\belowrulesep] 
&\multicolumn{1}{c}{(1)}&\multicolumn{1}{c}{(2)}&\multicolumn{1}{c}{(3)} \\[1.5ex]
& \textbf{Gender} & \textbf{Observations} & \textbf{Share} \tabularnewline 
\midrule
\textbf{Panel A. STEM} &&& \\
            &  Males    & 1722  & 0.92 \tabularnewline 
            &  Females  & 158   & 0.08 \tabularnewline
\midrule 
\textbf{Panel B. Social Sciences} &&& \\
&  Males    & 261   & 0.85 \tabularnewline
&  Females  &  47   & 0.15 \tabularnewline
\bottomrule 
\end{tabular}
}
     \caption{
        \textbf{University lecturers with habilitation in 1998.}\\
        \footnotesize{
        \emph{Notes.}- This table describes university lecturers with a habilitation---i.e., the pipeline for professors---in unified Germany in 1998 by subject group and by gender.  
        \emph{Sources.}- Personnel statistics for academic staff \citep{destatis1998personal}, academic staff with habilitation excluding professors, and own calculations.
        }
    }
    \label{tab:pipeline1}
\end{table}
\clearpage

\begin{table}[t!]
    \centering
    \resizebox{0.82\textwidth}{!}{%
        \begin{tabular}{l ccc}
\toprule
 & (1) & (2) & (3) \\
Criterion & Social Sc. & STEM & Difference \\
\midrule
\multicolumn{4}{l}{\textbf{Panel A. Research productivity}} \\
(Weighted) Publications per professor &  &  &  \\
\quad East &     12.62 &     13.11 &     -0.50 \\
\quad West &     17.81 &     17.21 &      0.59 \\
\quad East-West difference &     -5.19 &     -4.10 &     -1.09 \\
(Yearly) PhDs per professor &  &  &  \\
\quad East &      0.88 &      0.86 &      0.03 \\
\quad West &      1.30 &      1.26 &      0.04 \\
\quad East-West difference &     -0.41 &     -0.40 &     -0.01 \\
External funds per researcher (in 1000s) &  &  &  \\
\quad East &     23.86 &     27.30 &     -3.44 \\
\quad West &     23.10 &     44.60 &    -21.50 \\
\quad East-West difference &      0.76 &    -17.30 &     18.06 \\
\midrule
\multicolumn{4}{l}{\textbf{Panel B. Student evaluation}} \\
Overall degree &  &  &  \\
\quad East &      4.08 &      4.12 &     -0.04 \\
\quad West &      3.93 &      4.07 &     -0.14 \\
\quad East-West difference &      0.15 &      0.04 &      0.10 \\
Course offer &  &  &  \\
\quad East &      3.80 &      3.83 &     -0.03 \\
\quad West &      3.64 &      3.80 &     -0.16 \\
\quad East-West difference &      0.16 &      0.02 &      0.13 \\
Supervision &  &  &  \\
\quad East &      4.13 &      4.09 &      0.04 \\
\quad West &      3.98 &      4.04 &     -0.06 \\
\quad East-West difference &      0.15 &      0.05 &      0.10 \\
\midrule
Number of departments &  &  &  \\
\quad East &        27 &        19 &  \\
\quad West &       111 &        77 &  \\
\bottomrule
\end{tabular}

    }
    \caption{\textbf{\ac{CHE} University Ranking –- Means per subject group and region.}\\
    \footnotesize{
        \emph{Notes.}- This table shows averages from the \ac{CHE} University ranking by subject (Social Sciences in column~1, \ac{STEM} subjects in column~2) and by region (East in first line of each criterion, West in second line) for the years 2005-2007 (ranking cycle~III). Column~3 displays the difference of \ac{STEM} subjects and the Social Sciences. The last row for each criterion displays the East-West difference. \ac{STEM} subjects include computer science and medicine, while Social Sciences comprise business administration, economics, law and psychology. The criteria shown are the following: The publications per professor are defined as the number of publications in a department in a three-year period divided by the number of its professors. This is only available for business administration, economics and medicine. PhDs per professor are the number of completed PhD dissertations divided by the number of professors. External funds are calculated as the average amount (in 1000 EUR) of external funds spent over a three-year period divided by the average number of researchers, excluding those who are third-party funded, in that period. This is not available for law and medicine.
        Student evaluations are based on a survey where students evaluate (i) the degree overall, (ii) the course content and specialization, and (iii) availability and support of teaching staff on a scale from 1 (worst) to 5 (best).
        \emph{Sources.}- \ac{CHE} University Ranking \citep{che}, and own calculations.
    }}
    \label{tab:che_means}
\end{table}

\clearpage




\end{document}